\documentclass[preprint,showpacs,preprintnumbers,amsmath,amssymb]{revtex4}

\usepackage{graphicx}
\usepackage{dcolumn}
\usepackage{bm}
\usepackage{amsmath}
\usepackage{amssymb}
\usepackage{braket}

\usepackage{color}

\def\lesssim{\ \raise.3ex\hbox{$<$}\kern-0.8em\lower.7ex\hbox{$\sim$}\ }
\def\gesim{\ \raise.3ex\hbox{$>$}\kern-0.8em\lower.7ex\hbox{$\sim$}\ }

\begin{document}
\title{Superfluid properties of an ultracold Fermi gas with an orbital 
Feshbach resonance in the BCS-BEC crossover region}
\author{Taro Kamihori, Daichi Kagamihara, and Yoji Ohashi} 
\affiliation{Department of Physics, Keio University, 3-14-1 Hiyoshi, Kohoku-ku, Yokohama 223-8522, Japan} 
\begin{abstract}
We theoretically investigate superfluid properties of a two-band gas of $^{173}$Yb Fermi atoms with an orbital Feshbach resonance (OFR). To describe the BCS-BEC crossover region, we include superfluid fluctuations caused by inter-band and intra-band pairing interactions associated with OFR, by extending the strong-coupling theory developed by Nozi\`eres and Schmitt-Rink to the two-band case below the superfluid phase transition temperature; however, effects of an experimentally inaccessible deep bound state are removed, to model a real $^{173}$Yb Fermi gas near OFR. We show that the condensate fraction in the upper closed channel gradually becomes smaller than that in the lower open channel, as one moves from the strong- to the weak-coupling regime, because the OFR-pairing mechanism tunes the interaction strengths by adjusting the energy difference between the two bands. However, even when the closed-channel band is much higher in energy than the open-channel band in the weak-coupling regime, the magnitude of the superfluid order parameter in the closed channel is found to be still comparable to that in the open channel. As the reason for this, we point out a pair-tunneling effect by the OFR-induced inter-band interaction. Besides these superfluid quantities, we also examine collective modes, such as the Goldstone mode, Schmid (Higgs) mode, as well as Leggett mode, to clarify how they appear in the spectral weights of pair-correlation functions in each band. Since the realization of a multi-band superfluid Fermi gas is a crucial issue in cold Fermi gas physics, our results would contribute to the basic understanding of this type of Fermi superfluid in the BCS-BEC crossover region.
\end{abstract}
\maketitle
\par
\section{Introduction}
\par
Since the proposal of a Feshbach resonance using orbital degrees of freedom \cite{Zhang2015}, this so-called orbital Feshbach resonance (OFR) has attracted much attention as a candidate for the pairing mechanism of a $^{173}$Yb Fermi gas \cite{Xu2016,He2016,Cheng2016,Iskin2016,Iskin2017,Zhang2017,Wang2017,Mondal2018a,Mondal2018b,Deng2018,Zou2018,Yu2019,Klimin2019,Tajima2019,Laird2020}. Although the superfluid phase transition of this alkaline-earth-like Fermi gas has not been realized yet, experimental groups have succeeded in tuning the strength of a pairing interaction between $^{173}$Yb atoms, by adjusting the threshold energy of OFR \cite{Pagano2015,Hofer2015}. At present, this system can already be cooled down below the Fermi temperature $T_{\rm F}$ \cite{Fukuhara2007}. Furthermore, the lifetime of this system is relatively long [$\sim O(1~{\rm s})$] \cite{Zhang2015}. Thus, the realization of a superfluid $^{173}$Yb Fermi gas with OFR is very promising. Once the superfluid phase transition is achieved, using the tunable pairing interaction, we would be able to systematically examine superfluid properties of this system, from the weak-coupling BCS (Bardeen-Cooper-Schrieffer) regime to the strong-coupling BEC (Bose-Einstein condensation) regime.
\par
In cold Fermi gas physics, we have already had superfluid $^{40}$K and $^6$Li Fermi gases \cite{Jin2004,Zwierlein2004,Bartenstein2004,Kinast2004}, where a different kind of Feshbach resonance works as the pairing mechanism \cite{Chin2010} (which we call the magnetic Feshbach resonance (MFR) in this paper). However, the quest for a superfluid $^{173}$Yb Fermi gas is still important, because it enables us to examine {\it multi-band effects} in the superfluid phase. (Note that the superfluid $^6$Li and $^{40}$K Fermi gases belong to the {\it single-band} system.) Because the ultracold Fermi gas system is expected as a quantum simulator for the study of complicated quantum many-body phenomena discussed in other fields, having both single-band and two-band superfluid Fermi gases would be useful for such an application. 
\par
Here, we explain how OFR is different from MFR: MFR uses an active electron spin in the outer most $s$-orbital of alkaline-metal atoms, and the Zeeman effect associated with this electron spin plays an important role in tuning the interaction strength by adjusting an external magnetic field \cite{Chin2010}. In addition, the so-called broad Feshbach resonance is used in $^{40}$K and $^6$Li Fermi gases. In this type of resonance, although two bands called the open and closed channels take part in the Feshbach resonance, the latter band is much higher in energy than the former, as far as we consider the interesting BCS-BEC crossover region \cite{Levin2005,Bloch2008,Giorgini2008}. As a result, the closed channel actually only appears in the intermediate state of MFR \cite{Hulet2005}. Thus, one may examine the broad MFR case by using the ordinary {\it single-band} (open-channel) BCS model.
\par
In the ground state of a rare-earth $^{173}$Yb atom, since the outermost $s$-orbital is fully occupied, the MFR pairing mechanism does not work. Instead, OFR uses two orbital states, $^1 S_0$ and $^3 P_0$, in forming the open and closed channels. These channels are degenerate in the absence of an external magnetic field ($B=0$). This degeneracy is lifted by the {\it nuclear} Zeeman effect when $B\ne 0$, which is also used to tune the strength of a pairing interaction \cite{Zhang2015}. In addition, the resonance width of OFR is not so broad \cite{Zhang2015}. The energy difference between the two channels is at most the order of the Fermi energy $\varepsilon_{\rm F}$ in the BCS-BEC crossover region. Because of this small energy difference, the thermal occupation of the closed channel cannot be ignored. Thus, we need to treat this system as a {\it two-band} Fermi gas.
\par
In considering a $^{173}$Yb Fermi gas, previous work on two-band metallic superconductivity would be helpful: In 1959, Shul and co-workers extended the original BCS theory to the two-band case \cite{Shul1959}, and clarified that the behavior of the superfluid order parameters is sensitive to the inter-band pairing interaction. In 1966, Leggett predicted a collective mode (Leggett mode) being accompanied by a relative phase oscillation of two superconducting order parameters \cite{Leggett1966}. 
\par
Recently, these topics have also been discussed in the context of $^{173}$Yb Fermi gases: Within the mean-field BCS approximation, the temperature dependence of the two superfluid order parameters has been calculated, including the effects of an inter-band interaction \cite{Xu2016}. The possibility of the Leggett mode has been examined in Refs. \cite{Zhang2017,Klimin2019}; however, Ref. \cite{He2016} predicted that it is severely damped in a $^{173}$Yb Fermi gas.
\par
In this paper, we investigate the superfluid properties of a $^{173}$Yb Fermi gas in the BCS-BEC crossover region. To include pairing fluctuations associated with the OFR-induced tunable attractive interactions, we extend the strong-coupling theory developed by Nozi\`eres and Schmitt-Rink (NSR) \cite{NSR,Randeria,Engelbrecht} in the single-band case to the two-band system below the superfluid phase transition temperature $T_{\rm c}$. Following Ref. \cite{He2016}, we choose realistic values of scattering parameters of a $^{173}$Yb Fermi gas. In addition, as pointed out in Refs. \cite{Zhang2015,Xu2016,He2016,Mondal2018a,Mondal2018b}, in order to describe the current experimental situation for $^{173}$Yb Fermi gases, one needs to remove the {\it experimentally inaccessible} deep bound state from the theory \cite{Pagano2015,Hofer2015}. In this paper, this is achieved by extending a method proposed in the normal state \cite{Mondal2018a} to the superfluid phase below $T_{\rm c}$. We briefly note that the NSR scheme has been applied to examine the BCS-BEC crossover behavior of $T_{\rm c}$ \cite{Xu2016,Mondal2018a,Mondal2018b}, as well as the Leggett mode at $T=0$ \cite{He2016}, in a $^{173}$Yb Fermi gas. It has also been applied to the case with non-zero temperatures below $T_{\rm c}$, although a different parameter region from the $^{173}$Yb case is considered \cite{Klimin2019} (where the above-mentioned deep bound state is absent). 
\par
Within the NSR scheme, we consider (1) superfluid order parameters, (2) condensate fractions, and (3) superfluid collective modes [Goldstone mode \cite{Anderson1958}, Schmid (Higgs) mode \cite{Schmid1968,Schon1986}, and Leggett mode \cite{Leggett1966}]. We clarify the band dependence of these quantities in the whole BCS-BEC crossover region. For the Leggett mode, we confirm that it does not appear in a superfluid $^{173}$Yb Fermi gas \cite{He2016}.
\par
This paper is organized as follows: In Sec. II, we present our formulation based on NSR. We also explain how to remove the effects of the deep bound state from the theory there. In Sec. III, we show our numerical results on the superfluid order parameters, condensate fractions, and collective modes, in the BCS-BEC crossover region below $T_{\rm c}$. Throughout this paper, we set $\hbar=k_{\rm B}=1$, and the system volume is taken to be unity, for simplicity.
\par
\par
\section{Formulation}
\par
\par
\subsection{Model two-band superfluid Fermi gas}
\par
\par
To examine a $^{173}$Yb Fermi gas near OFR, we consider a four-component Fermi gas ((two bands)$\times$(two pseudo-spins)), described by the Hamiltonian \cite{Zhang2015,Xu2016,He2016,Zhang2017,Mondal2018a,Mondal2018b,Deng2018},
\begin{eqnarray}
H &=& \sum_{\bm p}\xi^{\rm o}_{\bm p}
\Bigl[c_{{\rm g},\downarrow,{\bm p}}^\dagger c_{{\rm g},\downarrow,{\bm p}} +         c_{{\rm e},\uparrow,{\bm p}}^\dagger   c_{{\rm e},\uparrow,{\bm p}}
\Bigr] 
+ \sum_{\bm p}\xi^{\rm c}_{\bm p}
\Bigl[c_{{\rm g},\uparrow,{\bm p}}^\dagger c_{{\rm g},\uparrow,{\bm p}} +
      c_{{\rm e},\downarrow,{\bm p}}^\dagger c_{{\rm e},\downarrow,{\bm p}}
\Bigr]
\nonumber \\
&+& U_0 \sum_{{\bm p},{\bm p}',{\bm q}}
\Bigl[c^\dagger_{{\rm e},\downarrow,{\bm p}+{\bm q}/2} 
      c^\dagger_{{\rm g},\uparrow,-{\bm p}+{\bm q}/2}
      c_{{\rm g},\uparrow,-{\bm p}'+{\bm q}/2} 
      c_{{\rm e},\downarrow,{\bm p}'+{\bm q}/2}
\nonumber\\
&{ }&~~~~~~~~~~~~~~~~~~~~~~~~~~~~~~~~~~~~~~~~~~~
     +c^\dagger_{{\rm e},\uparrow,{\bm p}+{\bm q}/2}
      c^\dagger_{{\rm g},\downarrow,-{\bm p}+{\bm q}/2}
      c_{{\rm g},\downarrow,-{\bm p}'+{\bm q}/2}
      c_{{\rm e},\uparrow,{\bm p}'+{\bm q}/2}
\Bigr]
\nonumber \\
&+& U_1 \sum_{{\bm p},{\bm p}',{\bm q}}   
\Bigl[c^\dagger_{{\rm e},\downarrow,{\bm p}+{\bm q}/2}
      c^\dagger_{{\rm g},\uparrow,-{\bm p}+{\bm q}/2}
      c_{{\rm g},\downarrow,-{\bm p}'+{\bm q}/2}
      c_{{\rm e},\uparrow,{\bm p}'+{\bm q}/2}
\nonumber \\
&{ }&~~~~~~~~~~~~~~~~~~~~~~~~~~~~~~~~~~~~~~~~~~~
     +c^\dagger_{{\rm e},\uparrow,{\bm p}+{\bm q}/2}
      c^\dagger_{{\rm g},\downarrow,-{\bm p}+{\bm q}/2}
      c_{{\rm g},\uparrow,-{\bm p}'+{\bm q}/2}
      c_{{\rm e},\downarrow,{\bm p}'+{\bm q}/2}
\Bigr].
\label{eq.1}
\end{eqnarray}
Here, $c_{\lambda,\sigma,{\bm p}}^\dagger$ is the creation operator of a $^{173}\mathrm{Yb}$ Fermi atom, where $\lambda={\rm g,e}$ represent two orbital states $^1 S_0$ and $^3 P_0$, respectively. $\sigma=\uparrow,\downarrow$ denote two nuclear spin states of an $I=5/2$ $^{173}$Yb atom, contributing to OFR. Among the four components, $|{\rm g},\downarrow\rangle$ and $|{\rm e},\uparrow\rangle$ form the open channel, with the kinetic energy $\xi^{\rm o}_{\bm p} = \varepsilon_{\bm p}-\mu={\bm p}^2/(2m)-\mu$, where $\mu$ is the Fermi chemical potential and $m$ is the mass of a $^{173}$Yb atom. The closed channel consists of $|{\rm g},\uparrow\rangle$ and $|{\rm e},\downarrow\rangle$, with the kinetic energy $\xi^{\rm c}_{\bm p} = \varepsilon_{\bm p}+\omega_{\rm th}/2-\mu$. Here, $\omega_{\rm th}/2$ is the energy difference between the two channels and $\omega_{\rm th}$ is referred to as the threshold energy of OFR. Experimentally, $\omega_{\rm th}/2$ is tunable by adjusting an external magnetic field \cite{Zhang2015}.
\par
In Eq. (\ref{eq.1}), $U_0~(<0)$ and $U_1$ represent an intra-band and inter-band pairing interaction, respectively. These are related to the observable scattering lengths $a_{s\pm}$ as \cite{Zhang2015,Xu2016,Zhang2017,Mondal2018a,Mondal2018b}
\begin{equation}
{4 \pi a_{s\pm} \over m}=
{U_\pm \over 
\displaystyle
1+U_\pm\sum_{\bm p}^{p_{\rm c}} {1 \over 2\varepsilon_{\bm p}}},
\label{eq.2}
\end{equation}
where $U_\pm=U_0\mp U_1$, and $p_{\rm c}$ is a high-momentum cutoff. For a $^{173}$Yb Fermi gas near OFR, we should take \cite{Pagano2015,Hofer2015}
\begin{eqnarray}
\begin{array}{l}
a_{s+}=1900a_0, \\
a_{s-}=200a_0,\\
\end{array}
\label{eq.2b}
\end{eqnarray}
where $a_0 = 0.529$\AA\, is the Bohr radius. For these realistic values ($a_{s+}/a_{s-}\gg 1$), Ref. \cite{He2016} points out that the Leggett mode does not appear. References \cite{Zhang2017,Klimin2019} take the different values, $k_{\rm F}a_{s+}=1$ and $a_{s+}/a_{s-}=0.8\simeq 1$ (where $k_{\rm F}$ is the Fermi momentum), and they show the existence of this mode. In this paper, although we mainly consider the $^{173}$Yb case in Eq. (\ref{eq.2b}), the latter case is also discussed.
\par
In the superfluid phase, it is convenient to rewrite Eq. (\ref{eq.1}) in the Nambu representation \cite{Schrieffer,Ohashi2003}:
\begin{eqnarray}
H =
\sum_{{\bm p},\alpha={\rm o,c}}
{\hat \Psi}_{\alpha,{\bm p}}^\dagger 
\left[
\xi^\alpha_{\bm p} \tau_3 + \Delta_\alpha \tau_1
\right] 
{\hat \Psi}_{\alpha,{\bm p}} 
+
{1 \over 4}
\sum_{{\bm q},\alpha,\alpha'={\rm o,c},j=1,2} 
U_{\alpha,\alpha'}{\hat \rho}^\alpha_j({\bm q})  
{\hat \rho}^{\alpha'}_j(-{\bm q})+E_0,
\label{eq.3}
\end{eqnarray}
where
\begin{equation}
E_0=
-{2|\Delta_+|^2 \over U_+}
-{2|\Delta_-|^2 \over U_-}
+\sum_{\bm{p},\alpha={\rm o,c}}\xi^\alpha_{\bm p},
\label{eq.4}
\end{equation}
and
\begin{eqnarray}
{\hat \Psi}_{{\rm o},{\bm p}}
=\left(
\begin{array}{c}
c_{{\rm e},\uparrow,{\bm p}} \\
c_{{\rm g},\downarrow,-{\bm p}}^\dagger
\end{array}
\right),
\label{eq.5}
\end{eqnarray}
\begin{eqnarray}
{\hat \Psi}_{{\rm c},{\bm p}}
=
\left(
\begin{array}{c}
c_{{\rm e},\downarrow,{\bm p}} \\
c_{{\rm g},\uparrow,-{\bm p}}^\dagger
\end{array}
\right),
\label{eq.6}
\end{eqnarray}
are the two-component Nambu fields in the open and closed channels,
 respectively. $\tau_i~(i=1,2,3)$ are the Pauli matrices acting on particle-hole space in each channel. In Eq. (\ref{eq.3}), 
\begin{equation}
\Delta_{\rm o} = 
U_0 \sum_{\bm p} 
\langle 
c_{{\rm g},\downarrow,{-\bm p}}
c_{{\rm e},\uparrow,{\bm p}}
\rangle 
+
U_1 \sum_{\bm p} 
\langle 
c_{{\rm g},\uparrow,-{\bm p}} 
c_{{\rm e},\downarrow,{\bm p}}
\rangle,
\label{eq.7}
\end{equation}
\begin{equation} 
\Delta_{\rm c} =
U_0 \sum_{\bm p} 
\langle 
c_{{\rm g},\uparrow,-{\bm p}} 
c_{{\rm e},\downarrow,{\bm p}}
\rangle 
+ 
U_1 \sum_{\bm p} 
\langle 
c_{{\rm g},\downarrow,{-\bm p}} 
c_{{\rm e},\uparrow,{\bm p}}
\rangle,
\label{eq.8}
\end{equation}
are the superfluid order parameters in the open and closed channels, respectively. In this paper, these are taken to be real and proportional to the $\tau_1$ component, without loss of generality. $\Delta_\pm$ in Eq. (\ref{eq.4}) are related to $\Delta_{\alpha={\rm o,c}}$ as $\Delta_\pm = \left[ \Delta_{\rm c} \mp \Delta_{\rm o} \right]/2$.
\par
The interaction term involved in Eq. (\ref{eq.3}) consists of the intra-band interaction ($U_{\rm o,o}=U_{\rm c,c}=U_0$), and the inter-band one ($U_{\rm o,c}=U_{\rm c,o}=U_1$). In this term,
\begin{eqnarray}
{\hat \rho}^\alpha_j ({\bm q}) &=& \sum_{\bm p} {\hat \Psi}_{ \alpha {\bm p}+{\bm q}/2}^\dagger \tau_j {\hat \Psi}_{ \alpha {\bm p}-{\bm q}/2}
\label{eq.9}
\end{eqnarray}
is the generalized density operator in the $\alpha$ channel \cite{Ohashi2003,Takada}. Since we are taking $\Delta_{\alpha={\rm o,c}}$ to be parallel to the $\tau_1$ component, ${\hat \rho}^\alpha_1$ and ${\hat \rho}^\alpha_2$ physically describe the amplitude and phase fluctuations of the superfluid order parameter in the $\alpha$ channel, respectively \cite{Ohashi2003}.
\par
\par
\begin{figure}[t]
\includegraphics[width=14cm]{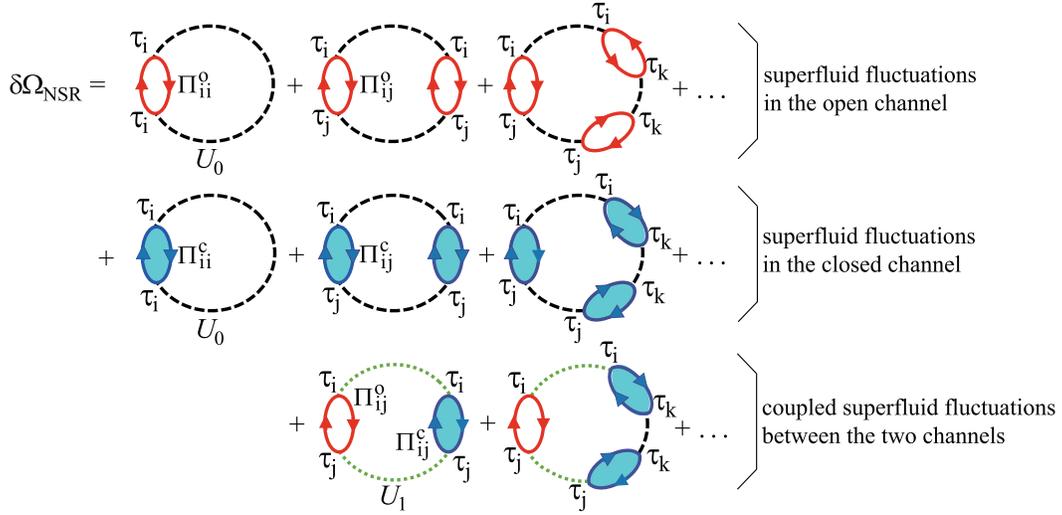}
\caption{NSR fluctuation corrections $\delta\Omega_{\rm NSR}$ to the thermodynamic potential $\Omega$ in the superfluid phase below $T_{\rm c}$. The diagrams in the first (second) line describe effects of superfluid fluctuations in the open (closed) channel, that are enhanced by the strong intra-band pairing interaction $U_0$. The diagrams in the third line involve, not only $U_0$, but also the inter-band interaction $U_1$, indicating that these diagrams physically describe effects of coupled superfluid fluctuations between the two channels. In each diagram, the bubbles are the pair correlation functions $\Pi_{ij}^{\alpha={\rm o,c}}$ ($i,j=1,2$) in Eq. (\ref{eq.13}): $\Pi_{11}^\alpha$ ($\Pi_{22}^\alpha$) physically describes amplitude (phase) fluctuations of the superfluid order parameter $\Delta_\alpha$ in the $\alpha$-channel. $\Pi_{12}^\alpha$ and $\Pi_{21}^\alpha$ represent the coupling of amplitude and phase fluctuations of the superfluid order parameter. $\tau_{i=1,2}$ are the Pauli matrices acting on particle-hole space in each channel.
}
\label{fig1}
\end{figure}
\par
\subsection{Amended NSR theory in the superfluid phase}
\par
\par
We include the effects of superfluid fluctuations in the BCS-BEC crossover region, by extending the NSR theory \cite{NSR} to the present model two-band system below $T_{\rm c}$. In this approach, the thermodynamic potential $\Omega=\Omega_{\rm MF}+\delta\Omega_{\rm NSR}$ consists of the mean-field BCS term $\Omega_{\rm MF}$ and the fluctuation correction $\delta\Omega_{\rm NSR}$. The former has the form,
\begin{eqnarray}
\Omega_{\rm MF}&=& -\frac{2|\Delta_+|^2}{U_+}  -\frac{2|\Delta_-|^2}{U_-} +\sum_{{\bm p},\alpha={\rm o,c}}
\Bigl[\xi^\alpha_{\bm p} -E^\alpha_{\bm p}
-2T\ln 
\left[
1+e^{-E^\alpha_{\bm p}/T}
\right]
\Bigr],
\label{eq.10}
\end{eqnarray}
where $E^{\alpha}_{\bm p}=\sqrt{{\xi^\alpha_{\bm p}}^2 + \Delta_\alpha^2}$ is the Bogoliubov single-particle dispersion in the $\alpha$ channel. The NSR correction term $\delta\Omega_{\rm NSR}$ involves the effects of superfluid fluctuations around the mean-field order parameters $\Delta_{\alpha={\rm o,c}}$, which is diagrammatically given in Fig. \ref{fig1}: In this figure, the first and second line describe the contribution of superfluid fluctuations in the open and closed channels by the intra-band interaction $U_0$, respectively. The last line describes the effects of channel-coupling by the inter-band interaction $U_1$. Summing up these diagrams, we have
\begin{eqnarray}
\delta \Omega_{\rm NSR}=
{T \over 2} \sum_{{\bm q},\nu_n} 
{\rm Tr}
\ln
\left[
1-{{\hat U} \over 2}{\hat \Pi}({\bm q},i\nu_n)
\right],
\label{eq.11}
\end{eqnarray}
where $\nu_n$ is the boson Matsubara frequency,
\begin{eqnarray}
{\hat U}
= 
\left(
\begin{array}{cccc}
U_0 & 0 &  U_1 & 0  \\
0 & U_0  & 0 &  U_1  \\
U_1 & 0 & U_0  & 0  \\
0 &  U_1 & 0 & U_0  \\
\end{array}
\right),
\label{eq.12}
\end{eqnarray}
and 
\begin{eqnarray}
{\hat \Pi}({\bm q},i\nu_n)=
\left(
\begin{array}{cccc}
\Pi^{\rm o}_{11}({\bm q},i\nu_n) &  
\Pi^{\rm o}_{12}({\bm q},i\nu_n)  & 0 & 0  \\
\Pi^{\rm o}_{21}({\bm q},i\nu_n) &  
\Pi^{\rm o}_{22}({\bm q},i\nu_n)  & 0 & 0  \\
0 & 0 & \Pi^{\rm c}_{11}({\bm q},i\nu_n)
 &  \Pi^{\rm c}_{12}({\bm q},i\nu_n) \\
0 & 0 & \Pi^{\rm c}_{21}({\bm q},i\nu_n)
 &  \Pi^{\rm c}_{22}({\bm q},i\nu_n) \\
\end{array}
\right).
\label{eq.13}
\end{eqnarray}
In Eq. (\ref{eq.13}),
\begin{eqnarray}
\Pi^\alpha_{ij}({\bm q},i\nu_n)&=& 
- \int_0^{1/T} d\tau  e^{i \nu_n \tau} \langle T_\tau [ \rho^\alpha_i ({\bm q},\tau)  \rho^\alpha_j (-{\bm q},0) ] \rangle
\nonumber \\ 
&=& T \sum_{{\bm p},\omega_m} \mathrm{Tr} 
\bigl[ 
\tau_i G^\alpha({\bm p}+{\bm q}/2, i\omega_m+i\nu_n)  \tau_j  G^\alpha({\bm p}-{\bm q}/2,i\omega_m)
\bigr]
\label{eq.14}
\end{eqnarray}
is the pair-correlation function, where $\omega_m$ is the fermion Matsubara frequency, and 
\begin{eqnarray}
\hat{G}^\alpha ({\bm p}, i \omega_m) = \frac{1}{i \omega_m -\xi^\alpha_{\bm p} \tau_3 -\Delta_\alpha \tau_1}
\label{eq.15}
\end{eqnarray}
is the $2\times 2$ matrix single-particle BCS Green's function \cite{Schrieffer}. The $\omega_m$-summation in Eq. (\ref{eq.14}) gives 
\begin{eqnarray}
\Pi^\alpha_{11}&=&\sum_{\bm p} \left[1- \frac{\xi^\alpha_{ {\bm p}+{\bm q}/2 }\xi^\alpha_{ {\bm p}-{\bm q}/2 }-\Delta_\alpha^2}{E^\alpha_{ {\bm p}+{\bm q}/2 } E^\alpha_{ {\bm p}-{\bm q}/2 }} \right]  \frac{E^\alpha_{ {\bm p}+{\bm q}/2 } -E^\alpha_{ {\bm p}-{\bm q}/2 }}{(E^\alpha_{ {\bm p}+{\bm q}/2 } -E^\alpha_{ {\bm p}-{\bm q}/2 })^2 + \nu_n^2} 
\left[ f(E^\alpha_{ {\bm p}+{\bm q}/2 } )-f(E^\alpha_{ {\bm p}-{\bm q}/2 } ) \right]
\nonumber \\
&-&\sum_{\bm p} \left[1+ \frac{\xi^\alpha_{ {\bm p}+{\bm q}/2 }\xi^\alpha_{ {\bm p}-{\bm q}/2 }-\Delta_\alpha^2}{E^\alpha_{ {\bm p}+{\bm q}/2 } E^\alpha_{ {\bm p}-{\bm q}/2 }} \right]  \frac{E^\alpha_{ {\bm p}+{\bm q}/2 } + E^\alpha_{ {\bm p}-{\bm q}/2 }}{(E^\alpha_{ {\bm p}+{\bm q}/2 } +E^\alpha_{ {\bm p}-{\bm q}/2 })^2 + \nu_n^2} 
\left[1-f(E^\alpha_{ {\bm p}+{\bm q}/2 } )-f(E^\alpha_{ {\bm p}-{\bm q}/2 } )\right],
\nonumber
\\
\label{eq.16}
\end{eqnarray}
\begin{eqnarray}
\Pi^\alpha_{22}&=&\sum_{\bm p} \left[1- \frac{\xi^\alpha_{ {\bm p}+{\bm q}/2 }\xi^\alpha_{ {\bm p}-{\bm q}/2 }+\Delta_\alpha^2}{E^\alpha_{ {\bm p}+{\bm q}/2 } E^\alpha_{ {\bm p}-{\bm q}/2 }} \right] \frac{E^\alpha_{ {\bm p}+{\bm q}/2 } -E^\alpha_{ {\bm p}-{\bm q}/2 }}{(E^\alpha_{ {\bm p}+{\bm q}/2 } -E^\alpha_{ {\bm p}-{\bm q}/2 })^2 + \nu_n^2}
\left[ f(E^\alpha_{ {\bm p}+{\bm q}/2 } )-f(E^\alpha_{ {\bm p}-{\bm q}/2 } ) \right]
\nonumber \\
&-&
\sum_{\bm p} \left[1+ \frac{\xi^\alpha_{ {\bm p}+{\bm q}/2 }\xi^\alpha_{ {\bm p}-{\bm q}/2 }+\Delta_\alpha^2}{E^\alpha_{ {\bm p}+{\bm q}/2 } E^\alpha_{ {\bm p}-{\bm q}/2 }} \right]  \frac{E^\alpha_{ {\bm p}+{\bm q}/2 } + E^\alpha_{ {\bm p}-{\bm q}/2 }}{(E^\alpha_{ {\bm p}+{\bm q}/2 } +E^\alpha_{ {\bm p}-{\bm q}/2 })^2 + \nu_n^2}
\left[ 1-f(E^\alpha_{ {\bm p}+{\bm q}/2 } )-f(E^\alpha_{ {\bm p}-{\bm q}/2 } ) \right],
\nonumber
\\
\label{eq.17}
\end{eqnarray}
\begin{eqnarray}
\Pi^\alpha_{12}&=&\sum_{\bm p} \left[ \frac{\xi^\alpha_{{\bm p}+{\bm q}/2}} {E^\alpha_{{\bm p}+{\bm q}/2}} -\frac{\xi^\alpha_{{\bm p}-{\bm q}/2}} {E^\alpha_{{\bm p}-{\bm q}/2}}  \right]  \frac{\nu_n}{(E^\alpha_{ {\bm p}+{\bm q}/2 } -E^\alpha_{ {\bm p}-{\bm q}/2 })^2 + \nu_n^2}
\left[ f(E^\alpha_{ {\bm p}+{\bm q}/2 } )-f(E^\alpha_{ {\bm p}-{\bm q}/2 } ) \right]
\nonumber
\\
&-&\sum_{\bm p} \left[ \frac{\xi^\alpha_{{\bm p}+{\bm q}/2}} {E^\alpha_{{\bm p}+{\bm q}/2}} +\frac{\xi^\alpha_{{\bm p}-{\bm q}/2}} {E^\alpha_{{\bm p}-{\bm q}/2}} \right]  \frac{\nu_n}{(E^\alpha_{ {\bm p}+{\bm q}/2 } +E^\alpha_{ {\bm p}-{\bm q}/2 })^2 + \nu_n^2}
\left[1-f(E^\alpha_{ {\bm p}+{\bm q}/2 } )-f(E^\alpha_{ {\bm p}-{\bm q}/2 } ) \right],
\nonumber
\\
\label{eq.18}
\end{eqnarray}
and $\Pi^\alpha_{21}=-\Pi^\alpha_{12}$. Here, $f(x)$ is the Fermi distribution function. Among these, $\Pi_{11}^\alpha$, $\Pi_{22}^\alpha$, and $\Pi_{12}^\alpha$, physically describe the amplitude and phase fluctuations of the superfluid order parameter, and the coupling between these fluctuations, respectively.
\par
In the NSR scheme, the superfluid order parameter $\Delta_\alpha$ in the $\alpha$ channel is determined from the extremum condition $\partial\Omega_{\rm MF}/\partial\Delta_\alpha=0$ for the mean-field thermodynamic potential $\Omega_{\rm MF}$ in Eq. (\ref{eq.10}). The resulting coupled gap equations are given by \cite{Zhang2015}
\begin{eqnarray}
{1 \over 2}
\left[
\frac{ \frac{\Delta_{\rm c}}{\Delta_{\rm o}} -1}{ \frac{4 \pi a_{s+}}{m} } - \frac{ \frac{\Delta_{\rm c}}{\Delta_{\rm o}} +1}{ \frac{4 \pi a_{s-}}{m} } 
\right]
=\sum_{\bm p} 
\left[
{1 \over 2E^{\rm o}_{\bm p}}
\tanh{E^{\rm o}_{\bm p} \over 2T} 
-{1 \over 2\varepsilon_{\bm p}}
\right],
\label{eq.19}
\end{eqnarray}
\begin{eqnarray}
{1 \over 2}
\left[
\frac{ \frac{\Delta_{\rm o}}{\Delta_{\rm c}} -1}{ \frac{4 \pi a_{s+}}{m} } - \frac{ \frac{\Delta_{\rm o}}{\Delta_{\rm c}} +1}{ \frac{4 \pi a_{s-}}{m} } 
\right]
=\sum_{\bm p} 
\left[
{1 \over 2E^{\rm c}_{\bm p}}
\tanh{E^{\rm c}_{\bm p} \over 2T} 
-{1 \over 2\varepsilon_{\bm p}}
\right],
\label{eq.20}
\end{eqnarray}
where the ultraviolet divergence has been absorbed into the scattering lengths $a_{s\pm}$ in Eq. (\ref{eq.2}). We briefly note that the gap equations (\ref{eq.19}) and (\ref{eq.20}) can also be obtained from Eqs. (\ref{eq.7}) and (\ref{eq.8}), when the expectation values in these equations are evaluated by using the BCS single-particle Green's functions ${\hat G}^{\alpha={\rm o,c}}({\bm p},i\omega_m)$ in Eq. (\ref{eq.15}).
\par
The gap equations (\ref{eq.19}) and (\ref{eq.20}) have two types of solutions \cite{He2016,Zhang2017,Iskin2016}: The {\it in-phase} solution [${\rm sgn}(\Delta_{\rm o})={\rm sgn}(\Delta_{\rm c})$], and the {\it out-of-phase} solution [${\rm sgn}(\Delta_{\rm o})=-{\rm sgn}(\Delta_{\rm c})$]. Between the two, the latter is known to be related to a shallow bound state \cite{Iskin2016,He2016} (see also Appendix A), and it is responsible for the recently observed OFR in a $^{173}$Yb Fermi gas \cite{Pagano2015,Hofer2015}. On the other hand, the in-phase solution is related to a deep bound state \cite{He2016}, and it has nothing to do with the observed OFR. This deep bound state is inaccessible in the current experiments on a $^{173}$Yb Fermi gas near OFR \cite{Pagano2015,Hofer2015}. Because we are interested in the observed situation, we focus only on the out-of-phase solution of Eqs. (\ref{eq.19}) and (\ref{eq.20}).
\par
Following the standard NSR approach \cite{Ohashi2003}, we solve the gap equations (\ref{eq.19}) and (\ref{eq.20}), together with the equation for the total number $N$ of Fermi atoms, to consistently determine $\Delta_{\rm o}$, $\Delta_{\rm c}$, and $\mu$. The number equation is obtained from the thermodynamic identity $N=-\partial \Omega/\partial \mu$ (where $\Omega=\Omega_{\rm MF}+\delta\Omega_{\rm NSR}$), giving
\begin{eqnarray}
N=\sum_{\alpha={\rm o,c}}N_0^\alpha
+{T \over 2}
\sum_{{\bm q},\nu_n}
{\rm Tr}
\left[
{\hat \Gamma}({\bm q},i\nu_n) 
{\partial \over \partial \mu}{\tilde \Pi}({\bm q},i\nu_n)
\right].
\label{eq.21}
\end{eqnarray}
Here, 
\begin{equation}
N_0^\alpha=\sum_{\bm p} 
\left[
1-{\xi^\alpha_{\bm p} \over E^\alpha_{\bm p}}
\tanh{E^\alpha_{\bm p} \over 2T} 
\right]
\label{eq.22}
\end{equation}
is the number equation in the mean-field level, and 
\begin{equation}
{\hat \Gamma}({\bm q},i\nu_n)=
\left[
1-{\tilde U}{\tilde \Pi}({\bm q},i\nu_n)
\right]^{-1}
\tilde{U}
\label{eq.23}
\end{equation}
is the $4\times 4$ particle-particle scattering matrix in the generalized random phase approximation (GRPA) \cite{Takada,Ohashi2003} in terms of the intra-band ($U_0$) and inter-band ($U_1$) interactions. In Eqs. (\ref{eq.21}) and (\ref{eq.23}), the ultraviolet divergence involved in the pair-correction function ${\hat \Pi}({\bm q},i\nu_n)$ in Eq. (\ref{eq.13}) has been removed as
\begin{eqnarray}
{\tilde \Pi}({\bm q},i\nu_n)=
\left(
\begin{array}{cccc}
{\tilde \Pi}^{\rm o}_{11}({\bm q},i\nu_n) &  
{\tilde \Pi}^{\rm o}_{12}({\bm q},i\nu_n)  & 0 & 0  \\
{\tilde \Pi}^{\rm o}_{21}({\bm q},i\nu_n) &  
{\tilde \Pi}^{\rm o}_{22}({\bm q},i\nu_n)  & 0 & 0  \\
0 & 0 & {\tilde \Pi}^{\rm c}_{11}({\bm q},i\nu_n)
 &  {\tilde \Pi}^{\rm c}_{12}({\bm q},i\nu_n) \\
0 & 0 & {\tilde \Pi}^{\rm c}_{21}({\bm q},i\nu_n)
 &  {\tilde \Pi}^{\rm c}_{22}({\bm q},i\nu_n) \\
\end{array}
\right)
=
{1 \over 2}{\hat \Pi}({\bm q},i\nu_n)+
\sum_{\bm p}{1 \over 2\varepsilon_{\bm p}}{\hat 1}.
\nonumber
\\
\label{eq.24}
\end{eqnarray}
Here, ${\hat 1}$ is the $4\times 4$ unit matrix. The interaction matrix ${\tilde U}$ in Eq. (\ref{eq.23}) has the form,
\begin{eqnarray}
{\tilde U}
=
{4\pi \over m}
\left(
\begin{array}{cccc}
a_{s0}& 0& a_{s1}& 0\\
0& a_{s0}& 0& a_{s1} \\
a_{s1}& 0& a_{s0}& 0\\
0& a_{s1}& 0& a_{s0}\\
\end{array}
\right),
\label{eq.25}
\end{eqnarray}
where 
\begin{eqnarray}
\begin{array}{l}
a_{s0}=[a_{s-}+a_{s+}]/2,\\
a_{s1}=[a_{s-}-a_{s+}]/2.\\
\end{array}
\label{as01}
\end{eqnarray}
We briefly note that the GRPA scattering matrix ${\hat \Gamma}({\bm q},i\nu_n)$ in Eq. (\ref{eq.23}) satisfies the required gapless condition for the Goldstone mode. That is, ${\hat \Gamma}({\bm q},i\nu_n)$ has a pole at ${\bm q}=\nu_n=0$, when the superfluid order parameters $\Delta_{\alpha={\rm o,c}}$ satisfy the coupled gap equations (\ref{eq.19}) and (\ref{eq.20}). For the proof, see Appendix B.
\par
Now, we remove the effects of the (experimentally inaccessible) deep bound state from the theory \cite{Xu2016,He2016,Mondal2018a,Mondal2018b}, in order to describe a real $^{173}$Yb Fermi gas near OFR \cite{Pagano2015,Hofer2015}. For this purpose, we extend the prescription developed in the normal state \cite{Mondal2018a,Mondal2018b} to the superfluid phase below $T_{\rm c}$: Noting that the particle-particle scattering matrix ${\hat \Gamma}({\bm q},i\nu_n)$ in Eq. (\ref{eq.23}) involves superfluid fluctuations associated with (i) the {\it shallow} bound state being responsible for OFR and (ii) the unwanted {\it deep} bound state, we first diagonalize ${\hat \Gamma}({\bm q},i\nu_n)$ as
\begin{eqnarray}
{\hat \Gamma}_{\rm d}({\bm q},i\nu_n)
\equiv 
{\hat W}^{-1} {\hat \Gamma}{\hat W}
=\left(
\begin{array}{cccc}
\Lambda^+_{11}({\bm q},i\nu_n) & 0 & 0 & 0  \\
0 &  \Lambda^+_{22}({\bm q},i\nu_n) & 0 & 0 \\
0 & 0 &  \Lambda^-_{11}({\bm q},i\nu_n) & 0   \\
0 & 0 & 0 &  \Lambda^-_{22}({\bm q},i\nu_n) \\       
\end{array}
\right), 
\label{eq.26}
\end{eqnarray}
where the $4\times 4$ matrix ${\hat W}$ diagonalizes ${\hat \Gamma}$. When $\omega_{\rm th}=0$ and $\nu_0=0$, the eigenvalues in Eq. (\ref{eq.26}) are reduced to
\begin{eqnarray}
\Lambda^\pm_{jj}({\bm q},0)
={4\pi a_{s\pm} \over m}
{1 \over \displaystyle
1-{4\pi a_{s\pm} \over m}{\tilde \Pi}_{jj}({\bm q},0)}.
\label{eq.27}
\end{eqnarray}
In Eq. (\ref{eq.27}), the scattering length $a_{s+}$ appears in $\Lambda^+_{jj}$. In addition, the pole equation $1-(4\pi a_{s+}/m){\tilde \Pi}_{22}(0,0)=0$ of $\Lambda^+_{22}({\bm q}=0,0)$ is just the same form as the gap equation in the {\it out-of-phase} case, given in Eq. (\ref{eq.a1b}) in Appendix A. Thus, one finds that $\Lambda^+_{11}$ and $\Lambda^+_{22}$, respectively, describe the amplitude and phase fluctuations of the superfluid order parameters associated with the shallow bound state \cite{Takada,Ohashi2003}. On the other hand, the pole condition $1-(4\pi a_{s-}/m){\tilde \Pi}_{22}(0,0)=0$ of $\Lambda^-_{22}({\bm q}=0,0)$ gives the gap equation (\ref{eq.a1}) in the {\it in-phase case}, so that $\Lambda^-_{jj}$ is found to be associated with the deep bound state. Thus, we remove the latter contribution by replacing the particle-particle scattering matrix ${\hat \Gamma}({\bm q},i\nu_n)$ in Eq. (\ref{eq.21}) with
\begin{eqnarray}
{\tilde \Gamma}({\bm q},i\nu_n)=
\left(
\begin{array}{cccc}
{\tilde \Gamma}_{\rm oo}^{11}  & 
{\tilde \Gamma}_{\rm oo}^{12}  &
{\tilde \Gamma}_{\rm oc}^{11}  & 
{\tilde \Gamma}_{\rm oc}^{12}  \\
{\tilde \Gamma}_{\rm oo}^{21}  & 
{\tilde \Gamma}_{\rm oo}^{22}  &
{\tilde \Gamma}_{\rm oc}^{21}  & 
{\tilde \Gamma}_{\rm oc}^{22}  \\
{\tilde \Gamma}_{\rm co}^{11}  & 
{\tilde \Gamma}_{\rm co}^{12}  &
{\tilde \Gamma}_{\rm cc}^{11}  & 
{\tilde \Gamma}_{\rm cc}^{12}  \\
{\tilde \Gamma}_{\rm co}^{21}  & 
{\tilde \Gamma}_{\rm co}^{22}  &
{\tilde \Gamma}_{\rm cc}^{21}  & 
{\tilde \Gamma}_{\rm cc}^{22}  \\
\end{array}
\right)
=
{\hat W}
\left(
\begin{array}{cccc}
\Lambda^+_{11}({\bm q},i\nu_m) & 0 & 0 & 0 \\
0 & \Lambda^+_{22}({\bm q},i\nu_m) & 0 & 0 \\
0 & 0 & 0 & 0 \\
0 & 0 & 0 & 0 \\
\end{array}
\right) 
{\hat W}^{-1}. 
\label{eq.28}
\end{eqnarray}
In Sec. III.A, we will show self-consistent solutions for the coupled gap equations (\ref{eq.19}) and (\ref{eq.20}) with the number equation (\ref{eq.21}) where ${\hat \Gamma}$ is replaced by ${\tilde \Gamma}$. 
\par
\par
\subsection{Condensate fraction $N_{\rm c}$}
\par
\par
The condensate fraction $N_{\rm c}=N_{\rm c}^{\rm o}+N_{\rm c}^{\rm c}$ physically describes the number of Bose-condensed Cooper pairs \cite{Yang}. Here,
\begin{equation}
N_{\rm c}^{\alpha={\rm o,c}}=
\sum_{\bm p}
|\langle
{\hat \Psi}^\dagger_{\alpha,{\bm p}}\tau_- {\hat \Psi}_{\alpha,{\bm p}}
\rangle|^2,
\label{eq.30}
\end{equation}
is the condensate fraction in the $\alpha$ channel \cite{Fukushima2007,Yang,Salasnich}, where $\tau_-=(\tau_1 - i\tau_2)/2$. Using Eq. (\ref{eq.15}), one obtains
\begin{eqnarray}
N_{\rm c}^\alpha=\sum_{\bm p}
\left|T\sum_{\omega_n}
{\rm Tr}\left[\tau_-{\hat G}^\alpha({\bm p},i\omega_m)\right]
\right|^2
=\sum_{\bm p}
{\Delta_\alpha^2 \over 4{E_{\bm p}^\alpha}^2}
\tanh^2 {E^\alpha_{\bm p} \over 2T}.
\label{eq.31}
\end{eqnarray}
\par
Equation (\ref{eq.31}) is consistent with the gap equations (\ref{eq.19}) and (\ref{eq.20}), in the sense that the latter equations can also be obtained from the same single-particle BCS Green's function ${\hat G}^\alpha$ in Eq. (\ref{eq.15}) as
\begin{equation}
\Delta_{\alpha={\rm o,c}}=U_0T\sum_{{\bm p},\omega_n}{\rm Tr}[\tau_-{\hat G}^\alpha({\bm p},i\omega_n)]+U_1T\sum_{{\bm p},\omega_n}{\rm Tr}[\tau_-{\hat G}^{-\alpha}({\bm p},i\omega_n)],
\label{eq.32}
\end{equation}
where ``$-\alpha$" means the opposite component to $\alpha$. On the other hand, ${\hat G}^\alpha$ in Eq. (\ref{eq.15}) cannot reproduce the whole expression for the number equation (\ref{eq.21}), but only gives the mean-field part $N_0^\alpha$. The same inconsistency also exists in the single-channel case \cite{NSR,Randeria,Engelbrecht}, where fluctuation corrections to the mean-field BCS single-particle Green's function are necessary, to obtain the NSR number equation. In the single-channel case, however, effects of these corrections on the condensate fraction are known to actually be very weak in the BCS-BEC crossover region \cite{Fukushima2007}. In this paper, therefore, we also examine the condensate fraction within Eq. (\ref{eq.31}).
\par
\par
\subsection{Collective excitations}
\par
\par
The energy of a collective mode associated with the superfluid order can be  determined from a pole of the analytic-continued particle-particle scattering matrix ${\tilde \Gamma}({\bm q},i\nu_n\to\omega+i\delta)$ in Eq. (\ref{eq.28}) (where $\delta$ is an infinitesimally small positive number). In this paper, we approximately evaluate the mode energy by only solving the real part of this pole condition \cite{Ohashi2003,Ohashi1997,note},
\begin{eqnarray}
{\rm Re} 
\left[ \det 
\left[
{\tilde \Gamma}^{-1} ({\bm q}, i\nu_n\to\omega+i\delta)
\right]
\right] =0.
\label{eq.33}
\end{eqnarray}
In particular, we set $\omega=c_{\rm s}q$ ($q\sim 0$) in looking for the gapless Goldstone mode with the sound velocity $c_{\rm s}$. The approximate mode equation (\ref{eq.33}) is valid for the case when the collective mode is weakly damped. To check this, we also examine how the spectral peak of this mode sharply appears in the spectral weights, given by
\begin{equation}
A_\alpha^{jj}({\bm q},\omega)=- 
{1 \over \pi}{\rm Im}
\left[
{\tilde \Gamma}_{\alpha\alpha}^{jj} ({\bm q}, i \nu_n\to\omega+i\delta)
\right],
\label{eq.34}
\end{equation}
where ${\tilde \Gamma}_{\alpha\alpha}^{jj}({\bm q},i\nu_n)~(\alpha={\rm o,c},~j=1,2)$ is given in Eq. (\ref{eq.28}). In (\ref{eq.34}), $A_\alpha^{11}$ and $A_\alpha^{22}$ are, respectively, the spectral weight of the amplitude correlation function of the superfluid order parameter and that of the phase correlation function in the $\alpha$ channel.
\par
\par
\section{Superfluid properties of a $^{173}$Yb Fermi gas}
\par
\begin{figure}
\includegraphics[width=6.5cm]{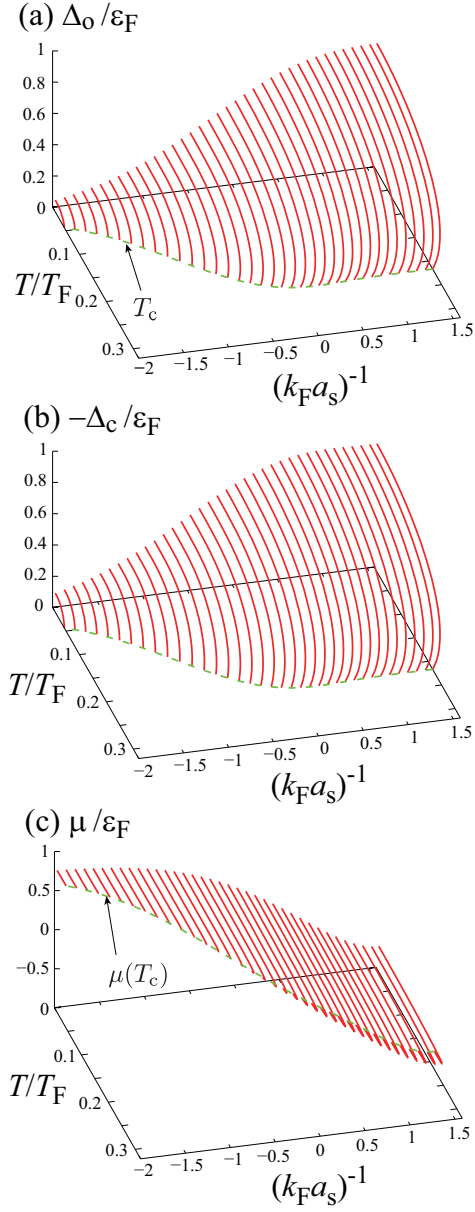}
\caption{Self-consistent out-of-phase solutions for the coupled gap equations (\ref{eq.19}) and (\ref{eq.20}) with the number equation (\ref{eq.21}) where ${\hat \Gamma}$ is amended as ${\tilde \Gamma}$ in Eq. (\ref{eq.28}). (a) Superfluid order parameter $\Delta_{\rm o}~(>0)$ in the open channel. (b) Superfluid order parameter $\Delta_{\rm c}~(<0)$ in the closed channel. (c) Fermi chemical potential $\mu$. $T_{\rm F}$, $\varepsilon_{\rm F}$, and $k_{\rm F}$ are, respectively, the Fermi temperature, Fermi energy, and Fermi momentum in an assumed single-band two-component free Fermi gas with the particle density $n=5 \times 10^{13}~{\rm cm}^{-3}$ \cite{Zhang2015,He2016,Xu2016}. The values of the scattering lengths $a_{s\pm}$ are given in Eq. (\ref{eq.2b}) \cite{Pagano2015,Hofer2015}, For these parameters, the threshold energy $\omega_{\rm th}$ vanishes when $(k_{\rm F}a_s)^{-1}=1.57$ \cite{Xu2016,Mondal2018a,Mondal2018b} (which experimentally corresponds to the vanishing external magnetic field). Thus, we only shows the results in the weaker coupling region, $(k_{\rm F}a_s)^{-1}\le 1.57$.}
\label{fig2}
\end{figure}
\par
\subsection{Superfluid order parameters $\Delta_\alpha$}
\par
Figure \ref{fig2} shows self-consistent (out-of-phase) solutions for the coupled gap equations (\ref{eq.19}) and (\ref{eq.20}) with the number equation (\ref{eq.21}) with ${\hat \Gamma}$ being replaced by ${\tilde \Gamma}$ in Eq. (\ref{eq.28}). The values of the scattering lengths $a_{s\pm}$ are given in Eq. (\ref{eq.2b}). In this figure, we measure the interaction strength in terms of $(k_{\rm F}a_s)^{-1}$ \cite{Zhang2015,Cheng2016,Xu2016,Mondal2018a,Mondal2018b}, where 
\begin{eqnarray}
\label{a_s}
a_s=
{a_{s0} - [a_{s0}^2 - a_{s1}^2]
\sqrt{m \omega_{\rm th}} \over 1-a_{s0} \sqrt{m \omega_{\rm th}}}
\label{eq.29}
\end{eqnarray}
is the $s$-wave scattering length in the open channel. $k_{\rm F}=[3\pi^2 n]^{1/3}$ is the Fermi momentum in an assumed single-band two-component Fermi gas with the particle density $n=5 \times 10^{13}~{\rm cm}^{-3}$ \cite{Zhang2015,He2016,Xu2016}. 
\par
Figures \ref{fig2}(a) and \ref{fig2}(b) indicate that the sign of $\Delta_{\rm c}$ is opposite to that of $\Delta_{\rm o}$, which is characteristic of the out-of-phase solution. In addition, one also finds from these figures that $|\Delta_{\rm o}|\simeq|\Delta_{\rm c}|$, and they exhibit a very similar temperature dependence to each other. As known in metallic superconductivity \cite{Shul1959,Ohashi2001}, these indicate the importance of the inter-band interaction $U_1$ in realizing a superfluid $^{173}$Yb Fermi gas. If $U_1$ is much weaker than the intra-band interaction $U_0$, their temperature dependence would be very different from each other near $T_{\rm c}$, especially in the weak-coupling BCS regime where the closed-channel band $\xi_{\bm p}^{\rm c}$ is much higher than the open-channel band $\xi_{\bm p}^{\rm o}$ \cite{Shul1959,Ohashi2001}. Thus, when the superfluid phase transition is achieved in a $^{173}$Yb Fermi gas, the observation of $|\Delta_{\alpha={\rm o,c}}|$ would provide useful information about the importance of the inter-band interaction.
\par
We see in Fig. \ref{fig2} that $\Delta_{\alpha={\rm o,c}}$ and $\mu$ exhibit weak first-order behavior near $T_{\rm c}$, when the interaction becomes strong to some extent. The same phenomenon also occurs in the single-band case, which is, however, known as an artifact of the NSR theory \cite{Fukushima2007}. This problem still exists in a more sophisticated strong-coupling theory, such as the self-consistent $T$-matrix approximation \cite{Haussmann2007}, and it is still unknown how to recover the expected second-order phase transition in the whole BCS-BEC crossover region. In this paper, therefore, leaving this problem as a future problem, we use the NSR data in Fig. \ref{fig2}, in calculating the condensate fraction $N_{\rm c}$, as well as the spectral weights, in the following discussions.
\par
\begin{figure}
\includegraphics[width=10cm]{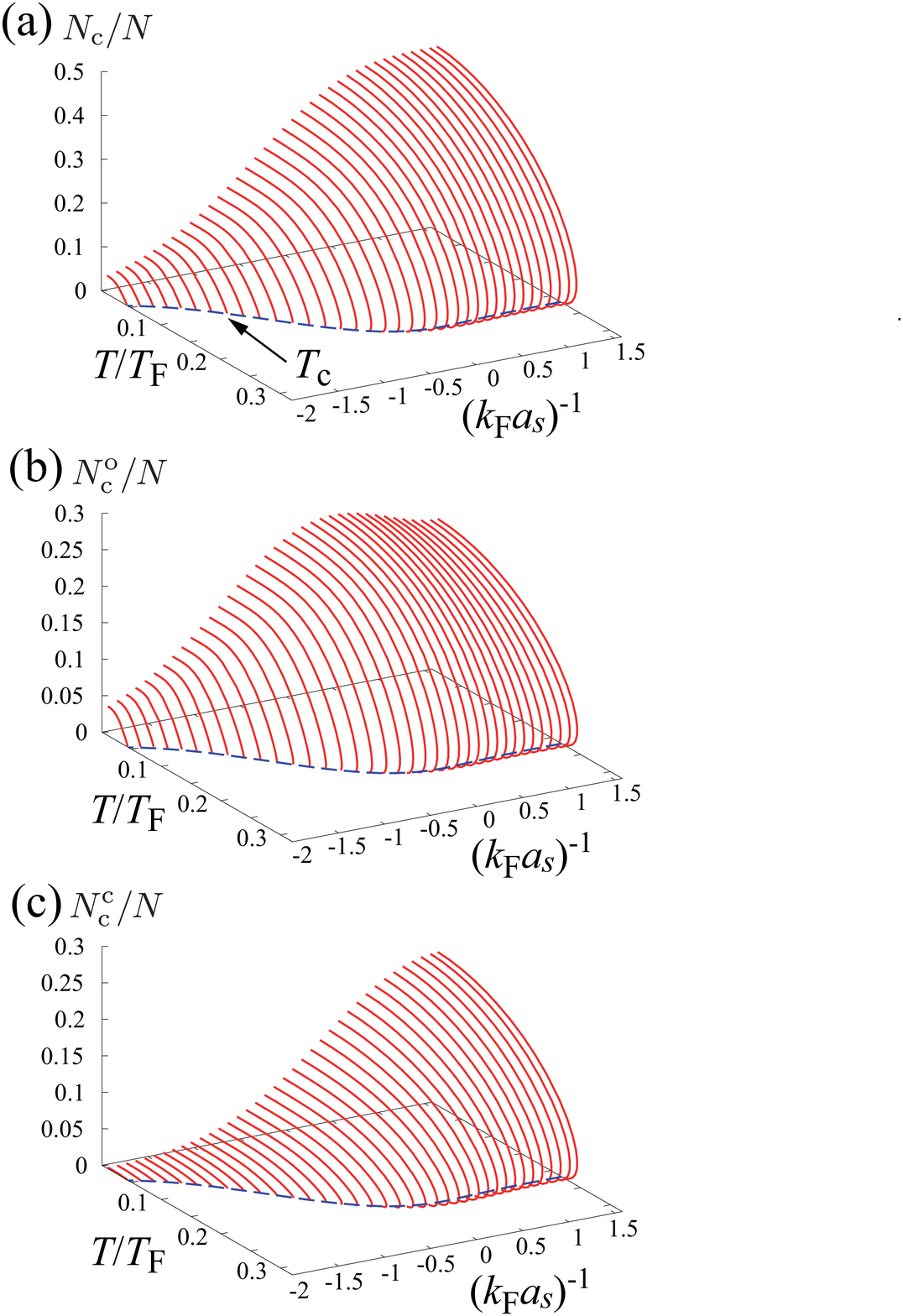}
\caption{(a) Calculated condensate fraction $N_{\rm c}$ in a superfluid $^{173}$Yb Fermi gas near OFR. Panels (b) and (c) show the open- and closed-channel components, respectively. In obtaining these results, we have used $\Delta_{\alpha={\rm o,c}}$ and $\mu$ in Fig. \ref{fig2}.
}
\label{fig3}
\end{figure}

\begin{figure}
\includegraphics[width=8cm]{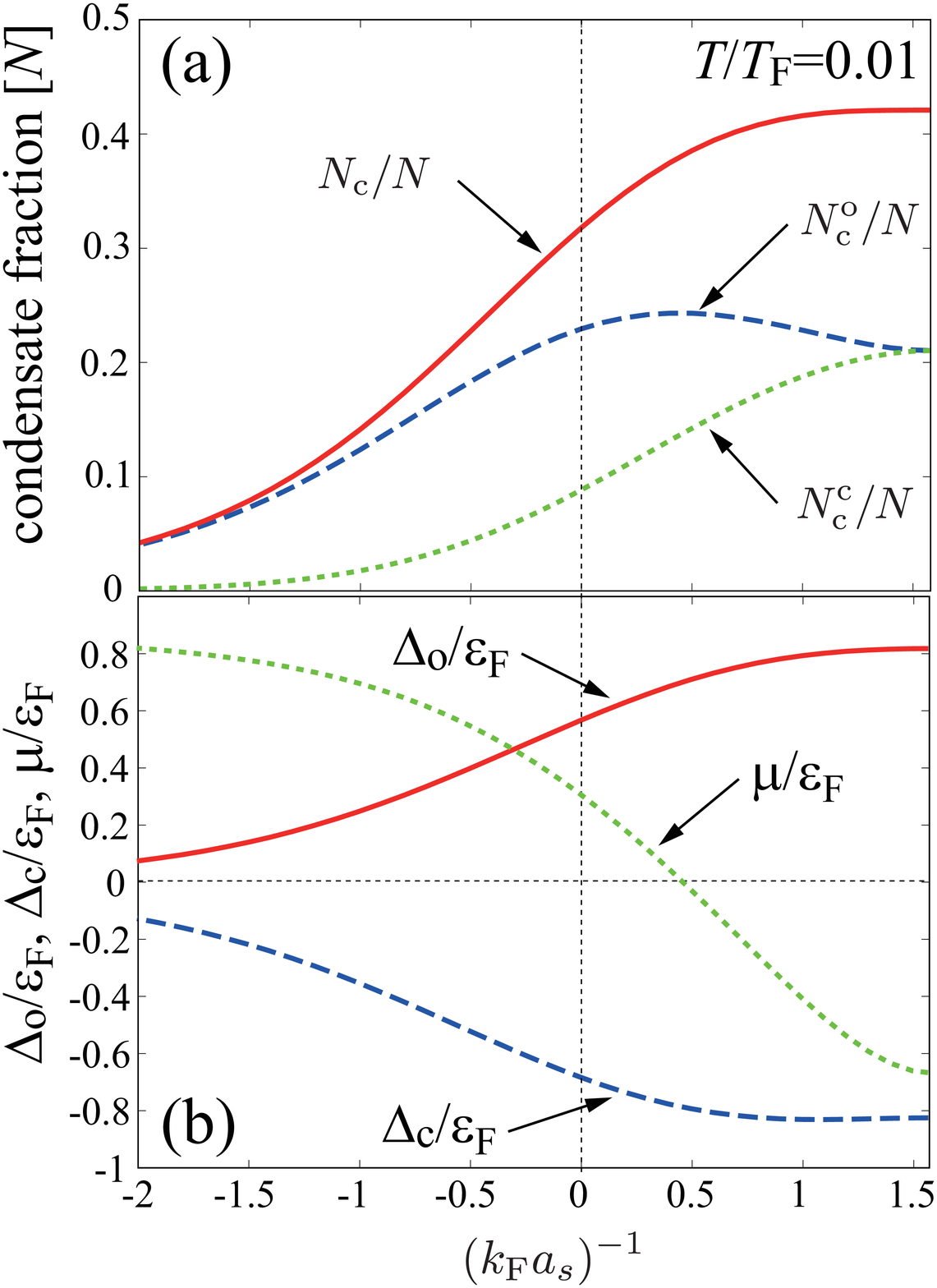}
\caption{(a) Condensate fractions $N_{\rm c}$ and $N_{\rm c}^{\alpha={\rm o,c}}$ as functions of the interaction strength $(k_{\rm F}a_s)^{-1}$. (b) Superfluid order parameters $\Delta_{\alpha={\rm o,c}}$ and Fermi chemical potential $\mu$ as functions of the interaction strength. We set $T/T_{\rm F}=0.01$.}
\label{fig4}
\end{figure}

\par
\subsection{Condensate fraction}
\par
Figure \ref{fig3} shows the condensate fraction in the BCS-BEC crossover regime of a two-band superfluid Fermi gas. For clarity, we also show in Fig. \ref{fig4}(a) the interaction dependence of this quantity far below $T_{\rm c}$. In the weak-coupling BCS regime ($(k_{\rm F}a_s)^{-1}\lesssim -1$), one sees in Fig. \ref{fig3}(a) that the total condensate fraction $N_{\rm c}$ is much smaller than $N/2$, which is realized when all the Fermi atoms form Cooper pairs. This is simply because only atoms near the Fermi surface contribute to the pair formation in this regime \cite{Fukushima2007,Salasnich}. 
\par
One also finds from Figs. \ref{fig3}(b), (c) and \ref{fig4}(a) that the condensate fraction $N_{\rm c}$ is dominated by the open channel component $N_{\rm c}^{\rm o}$ in the BCS regime. To understand this, we recall that the OFR-induced pairing interaction is tuned by adjusting the energy difference $\omega_{\rm th}/2$ between the upper closed channel and the lower open channel: The weak coupling regime is realized when $\omega_{\rm th}/2$ is large. As a result, most Fermi atoms occupy the open channel band in the weak-coupling regime, so that the number of atoms in the closed channel is very small. This naturally leads to $N_{\rm c}^{\rm c}\ll N_{\rm c}^{\rm o}\simeq N_{\rm c}$. Indeed, it has been shown that the closed channel is almost vacant, when $(k_{\rm F}a_s)^{-1}\lesssim -1$ \cite{Mondal2018a}. 
\par
In the weak-coupling BCS regime [$(k_{\rm F}a_s)^{-1}\lesssim -1$], in spite of $N_{\rm c}^{\rm c}\ll N_{\rm c}^{\rm o}$, the magnitude of the superfluid order parameter $|\Delta_{\rm c}|$ in the closed channel is still comparable to that in the open channel $|\Delta_{\rm o}|$ (see Figs. \ref{fig2}(a), \ref{fig2}(b) and \ref{fig4}(b)). More precisely, Fig. \ref{fig4}(b) shows that $|\Delta_{\rm c}|\gesim |\Delta_{\rm o}|$ (which is consistent with the previous work in the mean-field approximation \cite{Xu2016}). To consistently understand these different results ($N_{\rm c}^{\rm c}\ll N_{\rm c}^{\rm o}$ and $|\Delta_{\rm c}|\gesim |\Delta_{\rm o}|$), the key point is that, while the condensate fraction literally means the {\it number} of Bose-condensed Cooper pairs, the superfluid order parameter is related to the {\it binding energy} of a Cooper pair. Thus, although the number of Cooper pairs in the closed channel is very small in the weak-coupling BCS regime, each Bose-condensed Cooper pair in this channel has the {\it non-zero} binding energy $E_{\rm bind}^{\rm c}$, given by
\begin{eqnarray}
E_{\rm bind}^{\rm c}
=
\left\{
\begin{array}{ll}
2|\Delta_{\rm c}|&(\mu\ge\omega_{\rm th}/2),\\
2\left[\sqrt{(\omega_{\rm th}/2-\mu)^2+\Delta_{\rm c}^2} - (\omega_{\rm th}/2-\mu)\right]&(\mu<\omega_{\rm th}/2).\\
\end{array}
\right.
\label{eq.35}
\end{eqnarray}
In particular, deep inside the BCS regime (where $\omega_{\rm th}/2\gg\mu$), Eq. (\ref{eq.35}) gives $E_{\rm bind}^{\rm c}\simeq 2|\Delta_{\rm c}|\times(|\Delta_{\rm c}|/\omega_{\rm th})\ll 2|\Delta_{\rm c}|)$, which is much smaller than the binding energy $E_{\rm bind}^{\rm o}=2\Delta_{\rm o}$ in the open channel in this regime. This result may be interpreted as showing that the closed channel is in the {\it weaker} coupling regime than the open channel, which is consistent with the fact of a smaller condensate fraction $N_{\rm c}^{\rm c}\ll N_{\rm c}^{\rm o}$ in the former channel than that in the latter.
\par
The reason why $|\Delta_{\rm c}|\simeq |\Delta_{\rm o}|$ is obtained even in the weak-coupling BCS regime (where $N_{\rm c}^{\rm c}\ll N_{\rm c}^{\rm o}$) is that $\Delta_{\rm c}$ in Eq. (\ref{eq.8}) is made up of the pair amplitude, not only in the closed channel $\langle c_{{\rm g},\uparrow,-{\bm p}}c_{{\rm e},\downarrow,{\bm p}}\rangle$, but also in the open channel $\langle c_{{\rm g},\downarrow,-{\bm p}}c_{{\rm e},\uparrow,{\bm p}}\rangle$. Thus, even when the closed channel cannot produce the pair amplitude, $\Delta_{\rm c}$ can still become nonzero by using the pair amplitude supplied from the open channel through the inter-band interaction $U_1$. Regarding this, we note that the inter-band interaction in Eq. (\ref{eq.1}) may be viewed as a pair-tunneling term, where a pair of Fermi atoms moves from one band to the other.
\par
The total condensate fraction $N_{\rm c}$ increases monotonically upon increasing the interaction strength (see Fig. \ref{fig3}(a)). As shown in Figs. \ref{fig3}(c) and \ref{fig4}(a), the closed channel component $N_{\rm c}^{\rm c}$ also increases in this procedure, because the interaction strength is increased by decreasing the energy difference $\omega_{\rm th}/2$ between the two bands. The open- and closed-channel bands are degenerate at $(k_{\rm F}a_s)^{-1}=1.57$, at which $N_{\rm c}^{\rm c}=N_{\rm c}^{\rm o}$ is achieved. Although we cannot go beyond this coupling strength, Fig. \ref{fig4}(a) shows that more than 80\% of Fermi atoms already form Cooper pairs and are Bose-condensed far below $T_{\rm c}$ when $(k_{\rm F}a_s)^{-1}=1.57$.
\par
\begin{figure}[t]
\includegraphics[width=15cm]{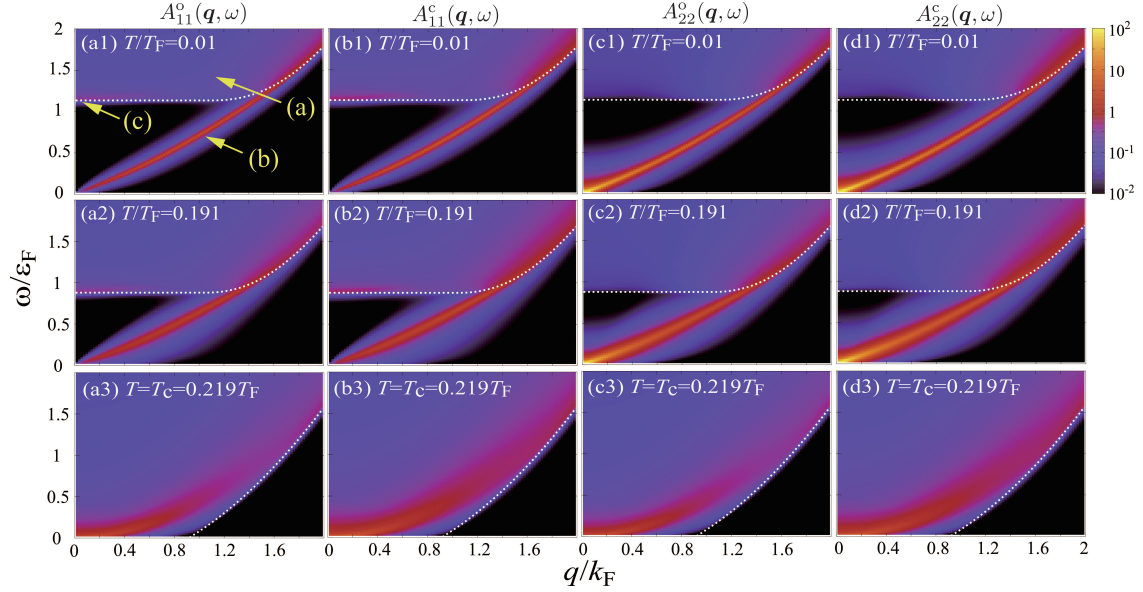}
\caption{Calculated spectral weights $A^{\alpha={\rm o,c}}_{jj}({\bm q},\omega)$ of the amplitude $(j=1)$ and phase ($j=2$) correlation functions in Eq. (\ref{eq.34}) in a unitary $^{173}$Yb superfluid Fermi gas ($(k_{\rm F}a_s)^{-1}=0$). In panel (a1), the continuum spectrum (a) describes two-particle excitations being accompanied by dissociation of a Cooper pair. The peak line (b) is the Goldstone mode. We also see a weak peak line (c), which is the Schmid (Higgs) mode \cite{Schmid1968,Schon1986}. The dotted line in each panel shows $E_{\rm th}^{\rm o}$ in Eq. (\ref{eq.36}), which gives the threshold energy of two-particle continuum. The spectral intensity is normalized by $2\pi^2/(mk_{\rm F})$. The same normalization is also used in Figs. \ref{fig6} and \ref{fig8}.
}
\label{fig5}
\end{figure}
\par
\subsection{Spectral weights and Goldstone mode}
\par
Figure \ref{fig5} shows that the spectral weights $A_{jj}^\alpha({\bm q},\omega)$ of the amplitude ($j=1$) and phase ($j=2$) correlation functions of the superfluid order parameters in a superfluid $^{173}$Yb Fermi gas at the unitarity. In this figure, we see that overall spectral structures are similar between the amplitude $(j=1)$ and phase $(j=2)$ components, as well as between the open ($\alpha={\rm o}$) and closed ($\alpha={\rm c}$) channels: They commonly have (a) the two-particle continuum, and (b) the peak line starting from ${\bm q}=\omega=0$ (see panel (a1)).
\par
Regarding the characteristic spectral structures (a) and (b), the two-particle continuum (a) is associated with the break-up of a Cooper pair. When the inter-band interaction $U_1$ is absent and the open-channel band is completely disconnected from the closed-channel band, they should have different threshold energies with respect to two-particle excitations, $E_{\rm th}^{\alpha={\rm o,c}}={\rm Min}[E^\alpha_{{\bm p}+{\bm q}/2}+E^\alpha_{-{\bm p}+{\bm q}/2}]$ \cite{Ohashi1997,Combescot2006}. That is,
\begin{eqnarray}
E_{\rm th}^{\rm o}=
\left\{
\begin{array}{ll}
2|\Delta_{\rm o}|& (\mu\ge 0,~{\rm and}~q\le2\sqrt{2m\mu}),\\
2\sqrt{[{q^2 \over 8m}-\mu]^2+|\Delta_{\rm o}|^2}& ({\rm otherwise}),
\end{array}
\right.
\label{eq.36}
\end{eqnarray}
\begin{eqnarray}
E_{\rm th}^{\rm c}=
\left\{
\begin{array}{ll}
2|\Delta_{\rm c}|& (\mu_{\rm c}\ge 0,~{\rm and}~q\le2\sqrt{2m\mu}),\\
2\sqrt{[{q^2 \over 8m}-\mu_{\rm c}]^2+|\Delta_{\rm c}|^2}& ({\rm otherwise}).
\end{array}
\right.
\label{eq.37}
\end{eqnarray}
Here, $\mu_{\rm c}=\mu-\omega_{\rm th}/2$. However, as seen in Fig. \ref{fig5},  the threshold energy $E_{\rm th}$ of two-particle excitations is commonly given by 
\begin{equation}
E_{\rm th}={\rm Min}[E_{\rm th}^{\rm o},E_{\rm th}^{\rm c}]=E_{\rm th}^{\rm o}.
\label{eq.Eth}
\end{equation}
This is because the inter-band interaction mixes the two channels, so that the depairing of Cooper pairs in the open channel also affects the two-particle continuum in the closed channel.
\par
\begin{figure}[t]
\includegraphics[width=11cm]{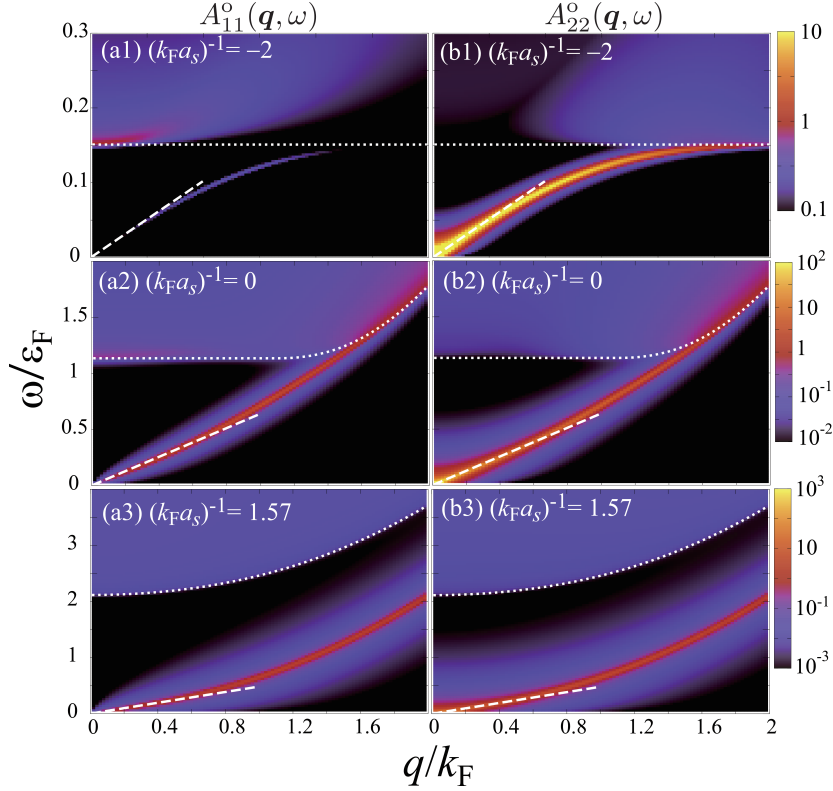}
\caption{Spectral weights $A^{\rm o}_{jj}({\bm q},\omega)$ ($j=1,2$) in the open channel. We set $T=0.01T_{\rm F}<T_{\rm c}$. (a1) and (b1): Weak-coupling BCS regime ($(k_{\rm F}a_s)^{-1}=-2$). (a2) and (b2): Unitary limit ($(k_{\rm F}a_s)^{-1}=0$). (a3) and (b3): Strong-coupling BEC regime ($(k_{\rm F}a_s)^{-1}=1.57$). The dotted line shows the threshold of two-particle continuum, given in Eq. (\ref{eq.36}). The dashed line shows the linear dispersion $\omega=c_{\rm s}q$ of the Goldstone mode, where the sound velocity $c_{\rm s}$ is determined from Eq. (\ref{eq.33}). Similar results are also obtained in the closed channel, although we do not explicitly show them here.}
\label{fig6}
\end{figure}
\par
In Fig. \ref{fig5}, the peak line (b) is just the (gapless) Goldstone mode in the superfluid phase, so that it becomes broad at $T_{\rm c}$ (see the lowest panels). Although this mode is physically interpreted as the collective {\it phase} oscillation of the superfluid order parameter, this spectral peak line actually appears in both $A_{11}^{\rm \alpha}({\bm q},\omega)$ (amplitude component) and $A_{22}^{\rm \alpha}({\bm q},\omega)$ (phase component). This is a result of the amplitude-phase coupling described by the correlation functions $\Pi_{12}^\alpha({\bm q},i\nu_n\to\omega+i\delta)$ in Eq. (\ref{eq.18}), as well as $\Pi_{21}^\alpha=-\Pi_{12}^\alpha$ \cite{Fukushima2007}. Because this coupling effect vanishes at $\omega=0$ (see Eq. (\ref{eq.18}) with $i\nu_n\to\omega+i\delta$), the peak intensity in $A_{11}^\alpha({\bm q},\omega)$ becomes weak around ${\bm q}=\omega=0$, as seen in the left two columns in Fig. \ref{fig5}. In addition, it is known that the amplitude-phase coupling also vanishes, when the system has the particle-hole symmetry with respect to the Fermi surface \cite{Takada,Ohashi1997,Randeria1997}. This situation is realized deep inside the BCS regime where the region near the Fermi surface dominantly contributes to the superfluid instability \cite{Takada,Ohashi1997,Randeria1997}. Indeed, as shown in Figs. \ref{fig6}(a1) and (a2), the peak intensity associated with the Goldstone mode is much weaker in $A_{11}^{\rm o}({\bm q},\omega)$ than in $A_{22}^{\rm o}({\bm q},\omega)$ in the BCS regime.
\par
\begin{figure}[t]
\includegraphics[width=8cm]{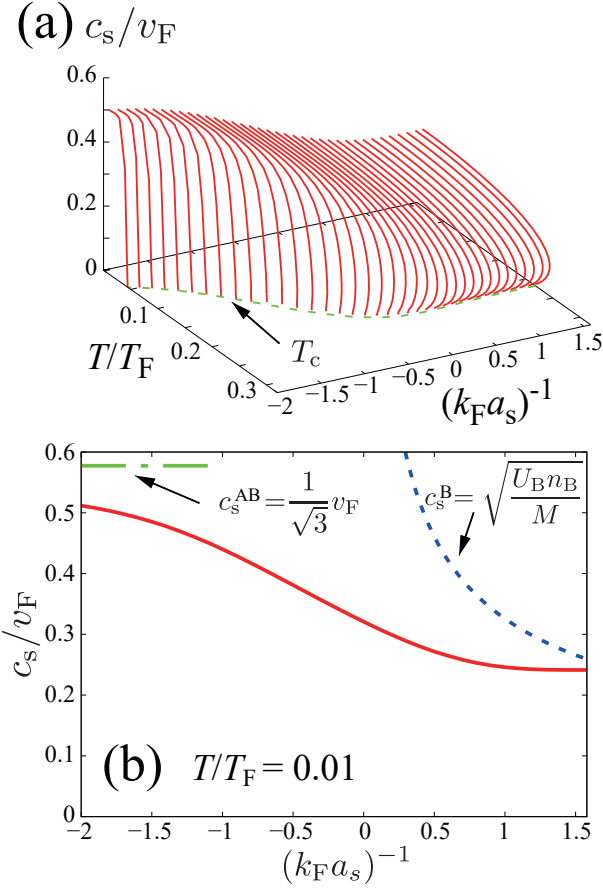}
\caption{(a) Calculated sound velocity $c_{\rm s}$ of the Goldstone mode in a $^{173}$Yb superfluid Fermi gas. $v_{\rm F}$ is the Fermi velocity of an assumed single channel Fermi gas with $N$ Fermi atoms. (b) $c_{\rm s}$ as a function of the interaction strength, when $T/T_{\rm F}=0.01$. $c_{\rm s}^{\rm AB}=v_{\rm F}/\sqrt{3}$ is the velocity of the Anderson-Bogoliubov mode in a single-band BCS superfluid. $c_{\rm s}^{\rm B}=\sqrt{U_{\rm B}n_{\rm B}/M}$ is the velocity of the  Bogoliubov phonon in an interacting Bose superfluid with $n_{\rm B}=N/2$ bosons with particle mass $M=2m$. $U_{\rm B}=4\pi a_{\rm B}/M$ is a repulsive interaction between bosons, where $a_{\rm B}=2a_s$ is the molecular scattering length. As mentioned in Sec. III.A, the first-order behavior seen in the strong-coupling regime is an artifact of the NSR theory.
}
\label{fig7}
\end{figure}
\par
In Fig. \ref{fig6}, the dashed line shows $\omega=c_{\rm s} q$, where the sound velocity $c_{\rm s}$ is evaluated from Eq. (\ref{eq.33}). The agreement of this linear dispersion with the spectral peak line confirms the validity of this approximate mode equation, at least in the present case. The same agreement is also obtained when we draw $\omega=c_{\rm s} q$ in all the panels in Fig. \ref{fig5} below $T_{\rm c}$, although we do not explicitly show the result here.
\par
We plot the calculated sound velocity $c_{\rm s}$ from Eq (\ref{eq.33}) in Fig. \ref{fig7}. In the weak-coupling BCS regime, $c_{\rm s}$ at low temperatures approaches the sound velocity of the Anderson-Bogoliubov mode \cite{Combescot2006,Anderson1958} in a {\it single-band} Fermi superfluid, given by
\begin{equation}
c_{\rm s}^{\rm AB}={v_{\rm F} \over \sqrt{3}},
\label{AB}
\end{equation}
with the Fermi velocity $v_{\rm F}=(3\pi^2 N)^{1/3}/m$ (see Fig. \ref{fig7}(b)). This is simply because most Fermi atoms occupy the open-channel band due to the fact that the upper closed-channel band is much higher than the lower open-channel band in this regime. Thus, the system properties in the BCS regime become close to the case of a single-band Fermi superfluid. 
\par
The two bands become degenerate ($\omega_{\rm th}/2=0$) when $(k_{\rm F}a_s)^{-1} = 1.57$. Thus, the system again may be treated as the single-band case there. Then, using the knowledge about the strong-coupling BEC regime of the ordinary single-band Fermi gas, the system in this regime is expected to be well described by a superfluid gas of $n_{\rm B}= N/2$ Bose molecules with a molecular mass $M=2m$ and a repulsive interaction $U_{\rm B}=4\pi a_{\rm B}/M$, where the molecular scattering length $a_{\rm B}$ equals $2a_s~(>0)$ within the NSR theory \cite{Randeria1997}. Indeed, Fig. \ref{fig7}(b) shows that, with increasing the interaction strength in the BEC side [$(k_{\rm F}a_s)^{-1}\gesim 0$], the sound velocity $c_{\rm s}$ approaches the velocity of the Bogoliubov phonon in such a Bose superfluid, given by
\begin{equation}
c_{\rm s}^{\rm B}=\sqrt{U_{\rm B}n_{\rm B} \over M}.
\label{BG}
\end{equation}
\par
The above discussions indicate that, even for the present two-band $^{173}$Yb superfluid Fermi gas, the character of the Goldstone mode still changes continuously from the Anderson-Bogoliubov mode in a {\it single-band} Fermi superfluid to the Bogoliubov phonon in the {\it ordinary} molecular Bose superfluid, with increasing the interaction strength. Thus, although the closed channel may affect $c_{\rm s}$ in the unitary regime [$(k_{\rm F}a_s)^{-1}\sim 0$] to some extent, the overall BCS-BEC crossover behavior of this quantity is similar to the single-band case \cite{Randeria2008}, such as $^{40}$K and $^6$Li superfluid Fermi gases \cite{Jin2004,Zwierlein2004,Bartenstein2004,Kinast2004}.
\par
\subsection{Absence of the Leggett mode in a $^{173}$Yb Fermi superfluid}
\par
Because the Leggett mode is the out-of-{\it phase} oscillation of the superfluid order parameters $\Delta_{\rm o}$ and $\Delta_{\rm c}$ \cite{Leggett1966}, it would appear in the {\it phase} component of the spectral weight $A_{22}^\alpha({\bm q},\omega)$, if it exists. However, one only sees the Goldstone mode (which is the in-phase oscillation of the superfluid order parameters) in Figs. \ref{fig5} and \ref{fig6}. This concludes the absence of the Leggett mode, at least in the low-energy region below the two-particle continuum. This conclusion agrees with the previous work by He and co-workers \cite{He2016}.
\par
\begin{figure}[t]
\includegraphics[width=8cm]{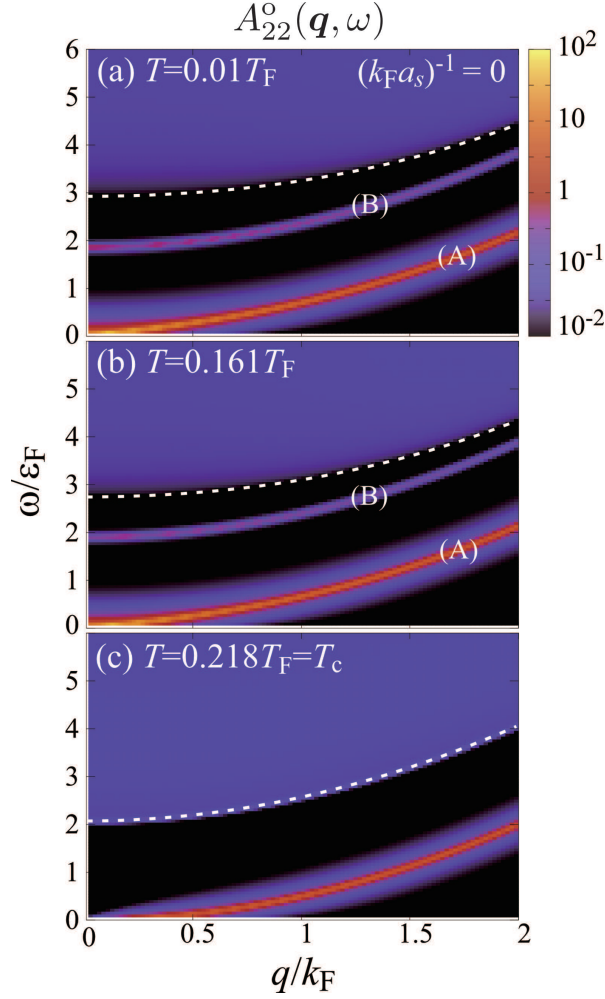}
\caption{Calculated phase component of the spectral weight $A^{\rm o}_{22}({\bm q},\omega)$, when $(k_{\rm F}a_s)^{-1}=0$. In this calculation, we take $a_{s+}=1900a_0$ and $a_{s-}=1520a_0$, and do not remove the deeper bound state. In the upper two panels ($T<T_{\rm c}$), the spectral peaks (A) and (B) correspond to the gapless Goldstone mode and the gapped Leggett mode, respectively. In panel (c), the low-energy peak line describes the dispersion of the non-condensed Bose molecules at $T_{\rm c}$. The dashed line shows the threshold of the two-particle continuum.
}
\label{fig8}
\end{figure}
\par
Recently, Refs. \cite{Zhang2017,Klimin2019} pointed out that the Leggett mode appears below the two-particle continuum, when we choose $a_{s-}/a_{s+}=0.8$. [Note that Eq. (\ref{eq.2b}) gives $a_{s-}/a_{s+}\simeq 0.11\ll 0.8$.] Indeed, considering this case by setting
\begin{eqnarray}
\begin{array}{l}
a_{s+}=1900a_0, \\
a_{s-}=1520a_0,\\
\end{array}
\label{eq.38}
\end{eqnarray}
we see in Fig. \ref{fig8} a gapped peak line (B) corresponding to the Leggett mode, in addition to the gapless peak line (A). In the case of Eq. (\ref{eq.38}), the binding energy $E_{{\rm bind}-}=1/(ma_{s-}^2)$ of the lower bound state is not so different from the binding energy $E_{{\rm bind}+}=1/(ma_{s+}^2)$ of the ``shallow" bound state as $E_{{\rm bind}-}/E_{{\rm bind}+}\simeq 1.5$. Thus, we have retained both bound states in obtaining Fig. \ref{fig8}. 
\par
\begin{figure}[t]
\includegraphics[width=8cm]{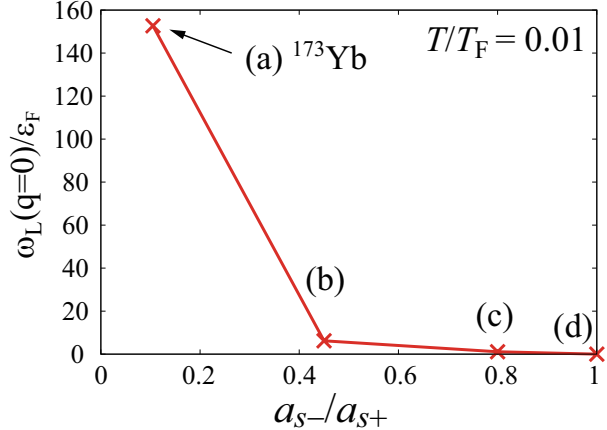}
\caption{Calculated energy $\omega_{\rm L}({\bm q}=0)$ of the Leggett mode as a function of the ratio $a_{s-}/a_{s+}$. We take $\omega_{\rm th}=0$, $T/T_{\rm F}=0.01$, and $a_{s+}=1900a_0$. (a) $a_{s-}=200a_0$ ($^{173}$Yb Fermi gas). (b) $a_{s-}=860a_0$  (c) $a_{s-}=1520a_0$ (Fig. \ref{fig8}). (d) $a_{s-}=1900a_0$.
}
\label{fig9}
\end{figure}

We find from Fig. \ref{fig9} that the energy of the Leggett mode increases with decreasing the ratio $a_{s-}/a_{s+}$ \cite{Zhang2017}. Thus, it seems that the present $^{173}$Yb Fermi gas near OFR [which has the scattering lengths in Eq. (\ref{eq.2b})] is not useful for the observation of the Leggett mode, even when the deep bound state becomes experimentally accessible. To observe this collective mode in the low-energy region ($\lesssim \varepsilon_{\rm F}$), we need to look for a different situation of a $^{173}$Yb Fermi gas, or another two-band Fermi gas with $a_{s-}\simeq a_{s+}$.
\par
\par
\begin{figure}[t]
\includegraphics[width=13cm]{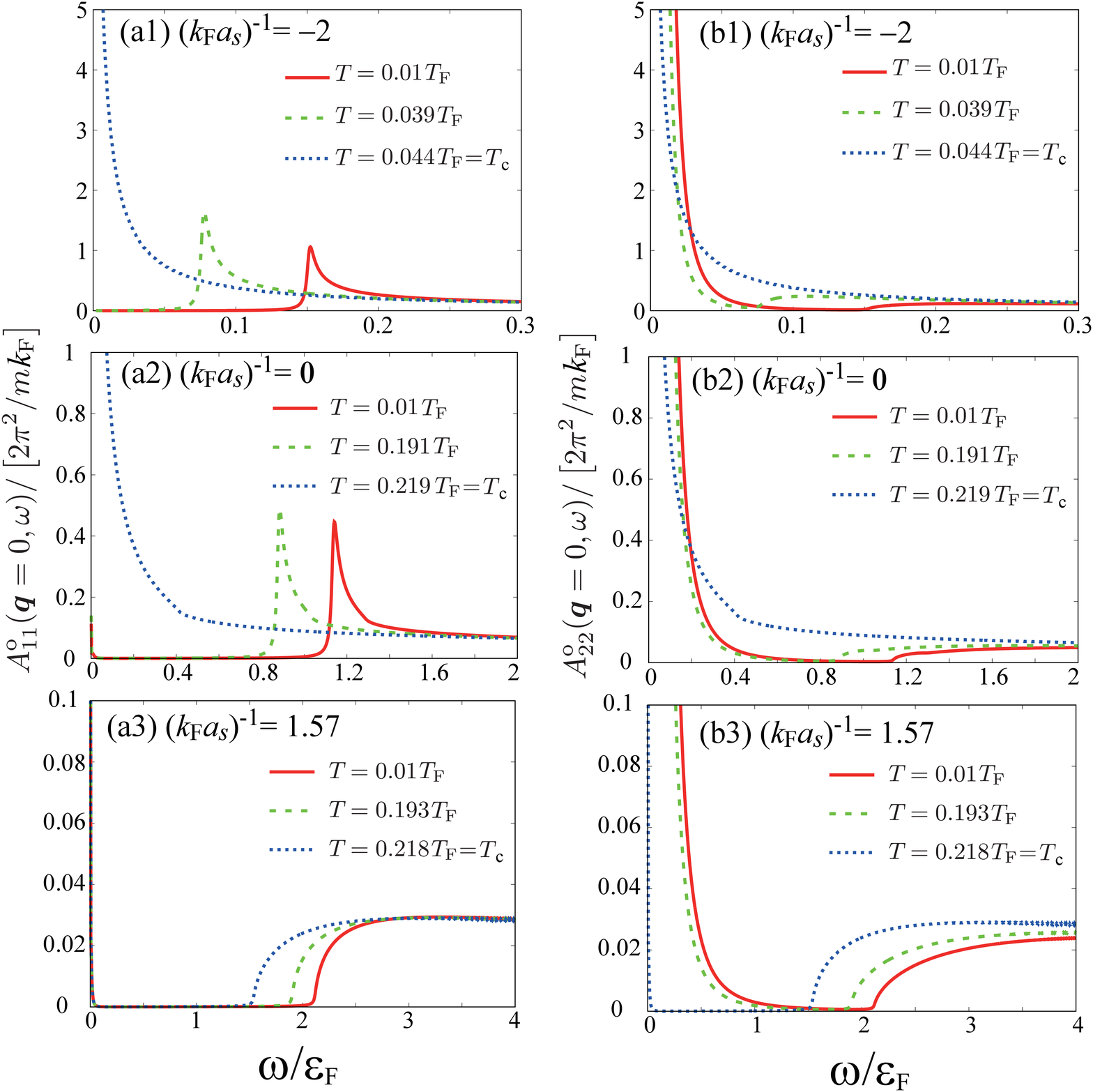}
\caption{Spectral weights $A^{\rm o}_{jj}({\bm q}=0,\omega)$ ($j=1,2$) as functions of $\omega$.}
\label{fig10}
\end{figure}
\par
\subsection{Schmid (Higgs) mode}
\par
The peak structure (c) in Fig. \ref{fig5}(a1) is the Schmid mode \cite{Schmid1968,Schon1986}, which is also referred to as the Higgs mode in the recent literature. The same peak is also seen in Figs. \ref{fig5}(a2), \ref{fig5}(b1), and \ref{fig5}(b2). This collective mode is accompanied by the amplitude oscillation of the superfluid order parameter, so that it appears in the amplitude component $A_{11}^{\alpha={\rm o,c}}({\bm q},\omega)$ of the spectral weight. Because of the amplitude-phase couplings $\Pi_{12}^\alpha$ and $\Pi_{21}^\alpha$, in principle, it may also appear in $A_{22}^\alpha({\bm q},\omega)$. However, because this collective mode appears at the threshold of two-particle continuum, it is not clearly seen in $A_{22}^\alpha({\bm q},\omega)$, as shown in the right two columns in Fig. \ref{fig5}. To confirm this more explicitly, we show in Fig. \ref{fig10} the energy dependence of $A_{jj}^{\rm o}({\bm q},\omega)$ at ${\bm q}=0$: This amplitude mode clearly appears as a peak structure at the threshold energy of the two-particle continuum in $A_{11}^{\rm o}({\bm q}=0,\omega)$, as shown in Figs. \ref{fig10}(a1) and (a2). However, such a peak structure is absent in the phase component $A_{22}^{\rm o}({\bm q}=0,\omega)$ shown in the right column in Fig. \ref{fig10}. 
\par
The Schmid (Higgs) mode is known to disappear in the BEC regime when $\mu<0$ \cite{Klimin2019,Pekker2015,Tempere2019,Ohashi2020}. Indeed, when $(k_{\rm F}a_s)^{-1}=1.57$ (where $\mu<0$, see Fig. \ref{fig4}(b)), the spectral peak no longer exists at the threshold of two-particle continuum in $A_{11}^{\rm o}({\bm q}=0,\omega)$, as shown in Fig. \ref{fig10}(a3). 
\par
When the inter-band interaction $U_1$ is absent and the open-channel band is disconnected from the closed-channel band, we would have two independent amplitude modes associated with $\Delta_{\rm o}$ and $\Delta_{\rm c}$. In the present case, however, since the Schmid mode appears at the threshold of the two-particle continuum, it is difficult to examine how the inter-band interaction causes the coupling of these two amplitude modes, which remains as our future problem. Regarding this, we point out that, in the field of metallic superconductivity, the Schmid mode is known to sometimes appear {\it below} the two-particle continuum \cite{Tsuchiya2013}. Thus, the realization of such a situation in an ultracold Fermi gas would be useful for the study of this problem.
\par
\par
\section{Summary}
\par
To summarize, we have discussed the superfluid properties of a $^{173}$Yb Fermi gas with an orbital Feshbach resonance (OFR). Including superfluid fluctuations within the framework of the strong-coupling theory developed by Nozi\`eres and Schmitt-Rink, and removing the effects of an experimentally inaccessible deep bound state, we self-consistently determined the superfluid order parameters in open and closed channels, as well as the Fermi chemical potential, as functions of temperature, in the BCS-BEC crossover region below $T_{\rm c}$. Using these data, we also calculated the condensate fractions, as well as the spectral weights of the amplitude and phase correlation functions of the superfluid order parameters in the two channels. From the spectral weights, we determined the sound velocity of the Goldstone mode in the crossover region. We also discussed the Leggett mode and Schmid (Higgs) mode.
\par
In the weak-coupling BCS regime, we showed that the condensate fraction in the closed channel ($N_{\rm c}^{\rm c}$) is much smaller than that in the open channel ($N_{\rm c}^{\rm o}$). This is because the OFR pairing mechanism tunes the interaction strength by adjusting the energy difference between the two channels: The weak-coupling BCS regime is realized when the closed-channel band is much higher than the open-channel band, so that the number of atoms, as well as the condensate fraction, in the closed channel becomes small there. 
\par
We also showed that the magnitude of the superfluid order parameter in the closed channel $|\Delta_{\rm c}|$ is comparable to that in the open channel $|\Delta_{\rm o}|$, even in the weak-coupling BCS regime. As the reason for this, we pointed out that the inter-band interaction $U_1$ plays a crucial role. Because the closed channel is almost vacant in the BCS regime, $\Delta_{\rm c}$ is dominantly made of the pair amplitude $\langle c_{{\rm g},\downarrow,-{\bm p}}c_{{\rm e},\uparrow,{\bm p}}\rangle$ produced in the open channel, which is transferred to the closed channel by the inter-band interaction.
\par
The closed-channel contribution to the condensate fraction increases with increasing the interaction strength, because the energy difference between the two bands becomes small. At the strongest interaction strength $(k_{\rm F}a_s)^{-1}=1.57$, they are degenerate, where $N_{\rm c}^{\rm c}=N_{\rm c}^{\rm o}$ is realized. Our result shows that more than 80\% of atoms take part in the total condensate fraction at this interaction strength.
\par
We have also examined the spectral weights $A_{jj}^{\alpha={\rm o,c}}({\bm q},\omega)$ of the amplitude ($j=1$) and phase ($j=2$) correlation functions of the superfluid order parameters. We found that, in both the open and closed channels, the threshold energy of the two-particle continuum is commonly determined by the superfluid order parameter $\Delta_{\rm o}$ in the open channel, as a result of the mixing of the two channels by the inter-band interaction. 
\par
Below the two-particle continuum, the gapless Goldstone mode appears as a spectral peak line starting from ${\bm q}=\omega=0$. The peak position was shown to be the same between the open-channel component $A_{jj}^{\rm o}({\bm q},\omega)$ and closed-channel one $A_{jj}^{\rm c}({\bm q},\omega)$. This means that a channel-selective measurement is not necessary in observing the Goldstone mode in a $^{173}$Yb superfluid Fermi gas. Regarding this, if the open and closed channels are disconnected from each other, these two channels would have different Goldstone modes with different sound velocities. In this sense, the fact that both channels have the same Goldstone mode also originates from the channel-mixing by the inter-band interaction.
\par
Besides the Goldstone mode, the Schmid (Higgs) mode was shown to appear at the threshold of the two-particle continuum in the amplitude component $A_{11}^{\alpha={\rm o,c}}({\bm q},\omega)$ of the spectral weight. However, the Leggett mode was not obtained below the two-particle continuum when the scattering lengths for a $^{173}$Yb Fermi gas with OFR are employed. This indicates that the present $^{173}$Yb Fermi gas with OFR may not be useful for the study of the Leggett mode. A proposal about how to observe this mode is a crucial theoretical issue in cold Fermi gas physics. Apart from this future problem, since the presence of an inter-band interaction is characteristic of multi-band Fermi gases, our results would be useful for the study of how this interaction affects the superfluid properties of a $^{173}$Yb Fermi gas in the BCS-BEC crossover region.
\par
\par
\begin{acknowledgments}
We thank D. Inotani, K. Manabe, and R. Sato for discussions. Y.O. was supported by a Grant-in-aid for Scientific Research from MEXT and JSPS in Japan (No.JP18K11345, No.JP18H05406, and No.JP19K03689).
\end{acknowledgments}
\par
\par
\appendix
\section{In-phase (out-of-phase) solution and deep (shallow) bound state}
\par
To see that the in-phase (out-of-phase) solution of the gap equations (\ref{eq.19}) and (\ref{eq.20}) is related to the deep (shallow) bound state, we consider the situation in which the open and closed channels are degenerate ($\omega_{\rm th}/2=0$) and $|\Delta_{\rm o}|=|\Delta_{\rm c}|$ is realized. In a $^{173}$Yb Fermi gas with OFR, it corresponds to the case when $(k_{\rm F}a_s)^{-1}=1.57$ \cite{Zhang2015,Xu2016}. For the in-phase solution ($\Delta_{\rm o}=\Delta_{\rm c}\equiv\Delta$), Eqs. (\ref{eq.19}) and (\ref{eq.20}) are reduced to the same BCS-type gap equation,
\begin{equation}
1=-{4\pi a_{s-} \over m}\sum_{\bm p}
\left[
{1 \over 2E_{\bm p}}\tanh{E_{\bm p} \over 2T}
-
{1 \over 2\varepsilon_{\bm p}}
\right],
\label{eq.a1}
\end{equation}
where $E_{\bm p}=\sqrt{(\varepsilon_{\bm p}-\mu)^2+\Delta^2}$. 
\par
For the out-of-phase solution ($\Delta_{\rm o}=-\Delta_{\rm c}\equiv\Delta$), one has
\begin{equation}
1=-{4\pi a_{s+} \over m}\sum_{\bm p}
\left[
{1 \over 2E_{\bm p}}\tanh{E_{\bm p} \over 2T}
-
{1 \over 2\varepsilon_{\bm p}}
\right].
\label{eq.a1b}
\end{equation}
Noting that $a_{s+}=1900a_0$ and $a_{s-}=200a_0$ in the case of a $^{173}$Yb Fermi gas near OFR \cite{Pagano2015,Hofer2015}, we solve the ``in-phase" gap equation (\ref{eq.a1}) in the strong-coupling regime (where $\mu<0$, $|\Delta/\mu|\ll 1$, and $|T/\mu|\ll 1$), which gives $\mu=-1/(2ma_{s-}^2)$. In this regime, on the other hand, the ``out-of-phase" gap equation (\ref{eq.a1b}) gives $\mu=-1/(2ma_{s+}^2)$. Since the Fermi chemical potential $\mu~(<0)$ approaches half the energy $E_{\rm bound}~(<0)$ of a two-body bound state in the BEC regime, one reaches,
\begin{eqnarray}
E_{\rm bound}\simeq
\left\{
\begin{array}{ll}
-\frac{1}{ma_{s-}^2}&({\rm in\mathchar`- phase}),\\
-\frac{1}{ma_{s+}^2}&({\rm out \mathchar`- of \mathchar`- phase}).\\
\end{array}
\right.
\label{eq.a2}
\end{eqnarray}
Together with the above-mentioned values of $a_{s\pm}$, we find from Eq. (\ref{eq.a2}) that the bound state in the in-phase case is much deeper than that in the out-of-phase case. Using the typical value of the number density $n=k_{\rm F}^3/(3\pi^2)=5\times 10^{13}~{\rm cm}^{-3}$ \cite{Zhang2015,He2016,Mondal2018a,Mondal2018b}, one has
\begin{eqnarray}
{|E_{\rm bound}| \over \varepsilon_{\rm F}}\simeq
\left\{
\begin{array}{ll}
138&({\rm in\mathchar`- phase}),\\
1.5&({\rm out \mathchar`- of \mathchar`- phase}).\\
\end{array}
\right.
\label{eq.a3}
\end{eqnarray} 
While $|E_{\rm bound}|$ is comparable to the Fermi energy $\varepsilon_{\rm F}$ in the out-of-phase case, it is much larger than $\varepsilon_{\rm F}$ in the in-phase case.
\par
\par
\section{Pole condition for ${\tilde \Gamma}(0,0)$}
\par
We prove that the particle-particle scattering matrix ${\tilde \Gamma}({\bm q},i\nu_n)$ in Eq. (\ref{eq.23}) has a pole at ${\bm q}=\nu_n=0$, when the gap equations (\ref{eq.19}) and (\ref{eq.20}) are satisfied. Noting that ${\tilde \Pi}^\alpha_{12}(0,0)=\Pi^\alpha_{21}(0,0)=0$ [see Eq. (\ref{eq.18})], one obtains the pole condition for ${\tilde \Gamma}(0,0)$ as,
\begin{eqnarray}
0&=&{\rm det}
\left[1-{\tilde U}{\tilde \Pi}(0,0)\right]
\nonumber
\\
&=&
{\rm det}
\left(
\begin{array}{cc}
1-{4\pi a_{s0} \over m}{\tilde \Pi}_{11}^{\rm o}(0,0) &
-{4\pi a_{s1} \over m}{\tilde \Pi}_{11}^{\rm c}(0,0) \\
-{4\pi a_{s1} \over m}{\tilde \Pi}_{11}^{\rm o}(0,0) &
1-{4\pi a_{s0} \over m}{\tilde \Pi}_{11}^{\rm c}(0,0) \\
\end{array}
\right)
\nonumber
\\
&\times&
{\rm det}
\left(
\begin{array}{cc}
1-{4\pi a_{s0} \over m}{\tilde \Pi}_{22}^{\rm o}(0,0) &
-{4\pi a_{s1} \over m}{\tilde \Pi}_{22}^{\rm c}(0,0) \\
-{4\pi a_{s1} \over m}{\tilde \Pi}_{22}^{\rm o}(0,0) &
1-{4\pi a_{s0} \over m}{\tilde \Pi}_{22}^{\rm c}(0,0) \\
\end{array}
\right),
\label{eq.b1}
\end{eqnarray}
where the scattering lengths $a_{s0}$ and $a_{s1}$ are given in Eq. (\ref{as01}). Because the superfluid order parameters $\Delta_{\alpha={\rm o,c}}$ are chosen to be parallel to the $\tau_1$ component in Eq. (\ref{eq.3}), the gapless Goldstone mode associated with phase fluctuations of the superfluid order parameters appears as the zero of the latter determinant in Eq. (\ref{eq.b1}). This pole condition can also be written as 
\begin{equation}
0={\rm det}[{\hat D}],
\label{eq.b2}
\end{equation}
where
\begin{eqnarray}
{\hat D}=
\left(
\begin{array}{cc}
\eta_+-{\tilde \Pi}_{22}^{\rm o}(0,0) & -\eta_-\\
-\eta_- & \eta_+-{\tilde \Pi}_{22}^{\rm c}(0,0)\\
\end{array}
\right).
\label{eq.b3}
\end{eqnarray}
Here, $\eta_\pm$ are related to the scattering lengths $a_{s\pm}$ in Eq. (\ref{eq.2b}) as,
\begin{equation}
\eta_\pm=
{1 \over 2}
\left[
{m \over 4\pi a_{s+}}\pm {m\over 4\pi a_{s-}}
\right].
\label{eq.b4}
\end{equation}
Summarizing the gap equations (\ref{eq.19}) and (\ref{eq.20}) as 
\begin{eqnarray}
0={\hat D}
\left(
\begin{array}{c}
\Delta_{\rm o} \\
\Delta_{\rm c} \\
\end{array}
\right),
\label{eq.b5}
\end{eqnarray}
we find that the pole equation (\ref{eq.b2}) is always satisfied, when the superfluid order parameters $\Delta_{\alpha={\rm o,c}}$ satisfy the gap equation (\ref{eq.b5}).
\par
\par

\end{document}